\documentclass[%
prc,
reprint,
superscriptaddress,
 amsmath,amssymb,
 aps,
]{revtex4-1}

\usepackage{color}
\usepackage{siunitx}
\usepackage{graphicx}
\usepackage{dcolumn}
\usepackage{bm}

\newcommand{\pA}    {p+A}
\newcommand{\dA}    {d+A}

\newcommand{\pdAu}  {p(d)+Au}
\newcommand{\AuAu}  {Au+Au}
\newcommand{\pPb}   {p+Pb}
\newcommand{\PbPb}  {Pb+Pb}
\newcommand{\Ru}	{^{96}_{44}\rm{Ru}}
\newcommand{\Zr}	{^{96}_{40}\rm{Zr}}
\newcommand{\RuRu}	{^{96}_{44}\rm{Ru} + ^{96}_{44}\rm{Ru}}
\newcommand{\ZrZr}	{^{96}_{40}\rm{Zr} + ^{96}_{40}\rm{Zr}}
\newcommand {\snn}  {\sqrt{s_{_{\rm NN}}}}
\newcommand{\sNN}{$\sqrt{s_{\rm NN}}$ }
\newcommand {\gevc} {GeV/$c$}
\newcommand{\gevcc}{GeV/$c^2$ }
\newcommand{ \Npart}{$N_{\rm part}$}
\newcommand {\pt}   {p_{T}}
\newcommand {\vexe} {v_{2,{\rm ebye}}}
\newcommand {\vnexe} {v_{n,{\rm ebye}}}
\newcommand {\psiPP}    {\psi_{\rm PP}}
\newcommand {\psiRP}    {\psi_{\rm RP}}
\newcommand {\psiEP}    {\psi_{\rm EP}}
\newcommand {\phires}   {\phi_{\rm res}}
\newcommand {\vres} {v_{2,{\rm reso}}}
\newcommand {\psiB}	{\psi_{B}}
\newcommand {\gSS}  {\gamma_{\rm SS}}
\newcommand {\gOS}  {\gamma_{\rm OS}}
\newcommand {\dg}	{\Delta\gamma}
\newcommand {\mult}	{N}
\newcommand {\minv}	{m_{\rm inv}}
\newcommand {\Bsq}	{B_{\rm sq}}

\newcommand {\mean}[1]  {\langle #1\rangle}

\begin{document}

\title{\textbf{Status of the Chiral Magnetic Effect Search in Relativistic Heavy-Ion Collisions}}

\author{ZHAO Jie}
\email{zhao656@purdue.edu}
\affiliation{College of Science, Huzhou University, Huzhou, Zhejiang 313000, China}
\affiliation{Department of Physics and Astronomy, Purdue University, West Lafayette, IN 47907, USA}
\author{TU Zhoudunming}
 \email{kongkong@rice.edu}
\affiliation{Department of Physics and Astronomy, Rice University, Houston TX 77054, USA}
\author{WANG Fuqiang}
\email{fqwang@purdue.edu}
\affiliation{College of Science, Huzhou University, Huzhou, Zhejiang 313000, China}
\affiliation{Department of Physics and Astronomy, Purdue University, West Lafayette, IN 47907, USA}

\date{\today}

\begin{abstract}
Quark interactions with topological gluon fields in QCD can yield local $\mathcal{P}$ and $\mathcal{CP}$ violations which could explain the matter-antimatter asymmetry in our universe. Effects of $\mathcal{P}$ and $\mathcal{CP}$ violations can result in charge separation under a strong magnetic field, a phenomenon called the chiral magnetic effect (CME). Experimental measurements of the CME-induced charge separation in heavy-ion collisions are dominated by physics backgrounds. Major theoretical and experimental efforts have been devoted to eliminating or reducing those backgrounds. We review the current status of these efforts in the search for the CME in heavy-ion collisions. 
\end{abstract}

\pacs{Valid PACS appear here}
\maketitle

\section{Introduction}
\label{sec:intro}
\vspace*{-1mm}

Quantum chromodynamics (QCD) governs the strong interaction among quarks and gluons.
Transitions between gluonic configurations from QCD vacuum fluctuations can be described by instantons/sphelarons and characterized by the Chern-Simons topological charge number~\cite{Lee:1973iz,Lee:1974ma,Morley:1983wr,Kharzeev:1998kz,Kharzeev:2004ey,Kharzeev:2007jp,Fukushima:2008xe,Kharzeev:2015znc}.
Quark interactions with gluonic fields, causing transitions of nonzero topological charges, would change their chirality (an imbalance in left- and right-handed quarks), leading to parity ($\mathcal{P}$) and charge conjugation parity ($\mathcal{CP}$) violations in local metastable domains~\cite{Kharzeev:1998kz,Kharzeev:2004ey,Kharzeev:2007jp,Fukushima:2008xe,Kharzeev:2015znc}. Such local $\mathcal{CP}$ violation in the strong interaction could explain the magnitude of the matter-antimatter asymmetry in the present universe~\cite{RevModPhys.76.1}.

In relativistic heavy-ion collisions, the approximate chiral symmetry is likely restored and the relevant degrees of freedom are quarks and gluons~\cite{Adams:2005dq,Adcox:2004mh,Arsene:2004fa,Back:2004je,Muller:2012zq}. In addition, an extremely strong magnetic field is produced by the spectator protons in the early times of those collisions~\cite{Kharzeev:2004ey,Kharzeev:2007jp,Fukushima:2008xe,Muller:2010jd,Kharzeev:2015znc}. It is possible that the magnetic field and the parity-violating local domains are on similar time scales in relativistic heavy-ion collisions. A chirality imbalanced domain of quarks under the strong magnetic field can then lead to a net electromagnetic current along the direction of the magnetic field~\cite{Kharzeev:2004ey,Kharzeev:2007jp,Fukushima:2008xe,Muller:2010jd,Kharzeev:2015znc}. This phenomenon is called the chiral magnetic effect (CME). 
Quarks hadronize into (charged) hadrons in the final state, leading to an experimentally observable charge separation.

An observation of the CME-induced charge separation in heavy-ion collisions would confirm several fundamental properties of QCD, namely, the approximate chiral symmetry restoration, topological charge fluctuations, and local $\mathcal{P}$ and $\mathcal{CP}$ violations. 
The measurements of such a charge separation would provide a means to study the non-trivial QCD topological structures in relativistic heavy-ion collisions~\cite{Lee:1973iz,Lee:1974ma,Morley:1983wr,Kharzeev:1998kz,Kharzeev:1999cz}.
Extensive theoretical efforts have been devoted to characterize the CME, and intensive experimental efforts have been invested to search for the CME in heavy-ion collisions at BNL's Relativistic Heavy Ion Collider (RHIC) and CERN's Large Hadron Collider (LHC)~\cite{Kharzeev:2015znc}.
\section{Early measurements and background contamination}
\label{sec:past}

In heavy-ion collisions, the particle azimuthal angle ($\phi$) distribution in momentum space is often described by a Fourier decomposition,
\begin{equation}
  \begin{split}
    \frac{dN}{d\phi} \propto  1& + 2v_{1}\cos(\phi-\psiRP) + 2v_{2}\cos2(\phi-\psiRP) + ...  \\
    & + 2a_{1}\sin(\phi-\psiRP) + 2a_{2}\sin2(\phi-\psiRP) + ... \,,
  \end{split}
  \label{eqThreeCtor3}
\end{equation}
where $\psiRP$ is the reaction-plane (RP) direction, 
defined to be the direction of the impact parameter vector and is expected on average to be perpendicular to the magnetic field direction. 
The parameters $v_{1}$ and $v_{2}$ account for the directed flow and elliptic flow~\cite{Reisdorf:1997flow}.
The parameters $a_{1,2}$  can be used to describe the charge separation effects. 
Usually only the first harmonic coefficient $a_{1}$ is considered.
Positively and negatively charged particles have opposite $a_{1}$ values, $a_{1}^{+}=-a_{1}^{-}$.
However, they average to zero  because of the random topological charge fluctuations from event to event~\cite{Kharzeev:2004ey}, 
making a direct observation of this parity violation effect impossible.
It is possible only via correlations, e.g.~measuring $\mean{a_{\alpha}a_{\beta}}$ with the average taken over all events in a given event sample.
The three-point $\gamma$ correlator is designed for this purpose~\cite{Voloshin:2004vk}, 
\begin{equation}
  \gamma = \mean{\cos(\phi_{\alpha}+\phi_{\beta}-2\psiRP)}\,.
  \label{eqThreeCtor4}
\end{equation}
Technically, the $\gamma$ correlator can also be calculated by the three-particle correlation method without an explicit determination of the RP~\cite{Voloshin:2004vk},
\begin{equation}
  \mean{\cos(\phi_{\alpha}+\phi_{\beta}-2\psiRP)}\approx\mean{\cos(\phi_{\alpha}+\phi_{\beta}-2\phi_c)}/v_{2,c}\,.
  \label{eqThreeCtor0}
\end{equation}
The role of the RP is instead fulfilled by the third particle, $c$, and $v_{2,c}$ is the elliptic flow parameter of the particle $c$.
The two sides in Eq.~(\ref{eqThreeCtor0}) would be equal if particle $c$ is correlated with particles $\alpha$ and $\beta$ via only the common correlation to the RP, without contamination of nonflow (few-particle) correlations between $c$ and $\alpha$ and/or $\beta$.

The $\gamma$ variable is vulnerable to particle correlation backgrounds, such as those caused by general momentum conservation~\cite{Pratt:2010zn,Bzdak:2010fd}. Those backgrounds are charge independent and thus the $\gamma$ difference between opposite-sign (OS) and same-sign (SS) charge pairs is usaully used to search for the CME,
\begin{equation}
    \dg = \gOS -\gSS\,.
  \label{eqThreeCtor5}
\end{equation}
Here OS ($+-$, $-+$) and SS ($++$, $--$) stand for the charge sign combinations of the $\alpha$ and $\beta$ particles.

\begin{figure}[htbp!]
  \centering 
  \includegraphics[width=0.4\textwidth]{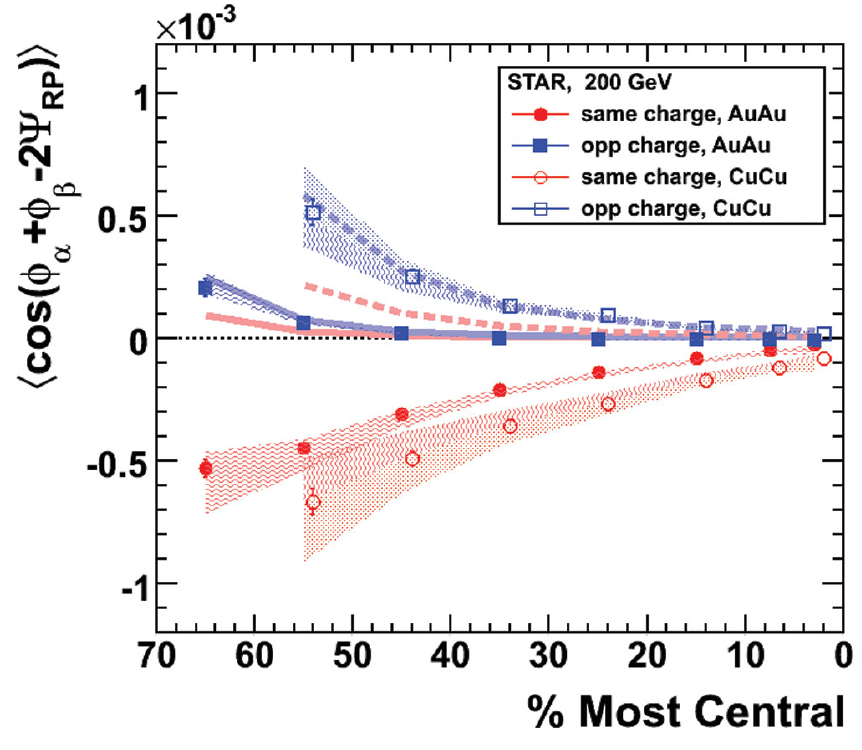}
  \caption{(Color online) The $\gamma$ correlators in Au+Au and Cu+Cu collisions at \sNN = 200 GeV by STAR~\cite{Abelev:2009ad}. Shaded bands represent uncertainty from the measurement of $v_{2}$. The thick solid (Au+Au) and dashed (Cu+Cu) lines represent HIJING calculations of the contributions from three-particle correlations. Collision centrality increases from left to right; 0\% corresponds to the most central collisions.}
  \label{FG_FirstCME}
\end{figure}

A significant $\dg$ has indeed been observed in heavy-ion collisions at RHIC and LHC~\cite{Abelev:2009ad,Abelev:2009ac,Adamczyk:2013hsi,Adamczyk:2014mzf,Abelev:2012pa,Ajitanand:2010}. 
Figure~\ref{FG_FirstCME} shows the $\gamma$ correlator as a function of the collision centrality in Au+Au and Cu+Cu collisions at \sNN = 200 GeV from STAR~\cite{Abelev:2009ad}.
Similarly, $\gOS$ and $\gSS$ correlators have been observed in Au+Au collisions at \sNN = 7.7-200 GeV from STAR~\cite{Adamczyk:2014mzf} and in Pb+Pb collisions at 2.76 TeV from ALICE~\cite{Abelev:2012pa}.
At high collision energies $\gOS$ is larger than $\gSS$, consistent with the CME expectations~\cite{Abelev:2009ad,Abelev:2009ac}. The difference between $\gOS$ and $\gSS$ decreases with increasing centrality, mainly because of the combinatorial dilution effect by the multiplicity. Under the CME scenario, such a decrease would also be consistent with the expectation of the magnetic field strength to decrease with increasing centrality~\cite{Kharzeev:2004ey,Kharzeev:2007jp,Fukushima:2008xe,Muller:2010jd,Kharzeev:2015znc}. 
At the low collision energy of \sNN =7.7 GeV, the difference between $\gOS$ and $\gSS$ disappears. This could be consistent with the disappearance of the CME at this energy, where hadronic interactions dominate~\cite{Adamczyk:2014mzf}.
Thus, the $\gamma$ correlator measurements are qualitatively consistent with the CME expectation~\cite{Abelev:2009ad,Abelev:2009ac,Adamczyk:2013hsi,Adamczyk:2014mzf}.

There are, however, mundane physics that could produce the same effect as the CME in the $\dg$ variable~\cite{Wang:2009kd,Pratt:2010zn,Bzdak:2010fd,Zhao:2018ixy}.
An example would be decays of resonances (or clusters in general) coupled with their $v_{2}$~\cite{Wang:2009kd,Wang:2016iov}; the $\dg$ variable is ambiguous between a back-to-back OS pair from the CME perpendicular to the RP and an OS pair from a resonance decay along the RP.
The resonance background was pointed out earlier but the magnitude estimate of the background contribution was wrong by 1-2 orders of magnitude~\cite{Voloshin:2004vk}.
Calculations with local charge conservation and momentum conservation effects can almost fully account for the measured $\dg$ signal at RHIC~\cite{Pratt:2010zn,Schlichting:2010qia,Bzdak:2010fd}. 
A Multi-Phase Transport (AMPT)~\cite{Zhang:1999bd,Lin:2001zk,Lin:2004en} model simulations can also largely account for the measured $\dg$ signal~\cite{Ma:2011uma,Shou:2014zsa}.
In general, these backgrounds are generated by two particle correlations (e.g.~from resonance decays) coupled with elliptic flow of the parent sources (resonances):
\begin{equation}
  \mean{\cos(\phi_{\alpha}+\phi_{\beta}-2\psi_{RP})} \approx \mean{\cos(\phi_{\alpha}+\phi_{\beta}-2\phires}\cdot \vres\,,
  \label{EQ_2}
\end{equation}
where $\mean{\cos(\alpha+\beta-2\phires)}$ is the angular correlation from the resonance decay, $\vres$ is the $v_{2}$ of the resonance.
The factorization of $\mean{\cos(\alpha+\beta-2\phires)}$ with $\vres$ is only approximate, because both depend on $\pt$ of the resonance~\cite{Wang:2016iov}.

The first unambiguous experimental evidence that background dominates was from small system collisoins~\cite{Khachatryan:2016got}.
The small system p+A or d+A collisions provide a control experiment, where the CME signal can be ``turned off'', but the $v_{2}$-related backgrounds still persist.
In non-central heavy-ion collisions, the $\psiPP$, although fluctuating~\cite{Alver:2006wh}, is generally aligned with the RP, thus generally perpendicular to the magnetic field. 
The $\dg$ measurement is thus $\emph{entangled}$ by the two contributions of the possible CME and the $v_2$-induced background.
In small-system p+A or d+A collisions, however, the $\psiPP$ is determined purely by geometry fluctuations, uncorrelated to the impact parameter or the magnetic field direction~\cite{Khachatryan:2016got,Belmont:2016oqp,Tu:2017kfa}. 
As a result, any CME signal would average to zero in the $\dg$ measurements with respect to the $\psiPP$.
Background sources, on the other hand, contribute to small-system p+A or d+A collisions similarly as to heavy-ion collisions.
Comparing the small system p+A or d+A collisions to A + A collisions could thus further our understanding of the background issue in the $\dg$ measurements.

\begin{figure}[htbp!]
  \centering 
  \includegraphics[width=0.40\textwidth]{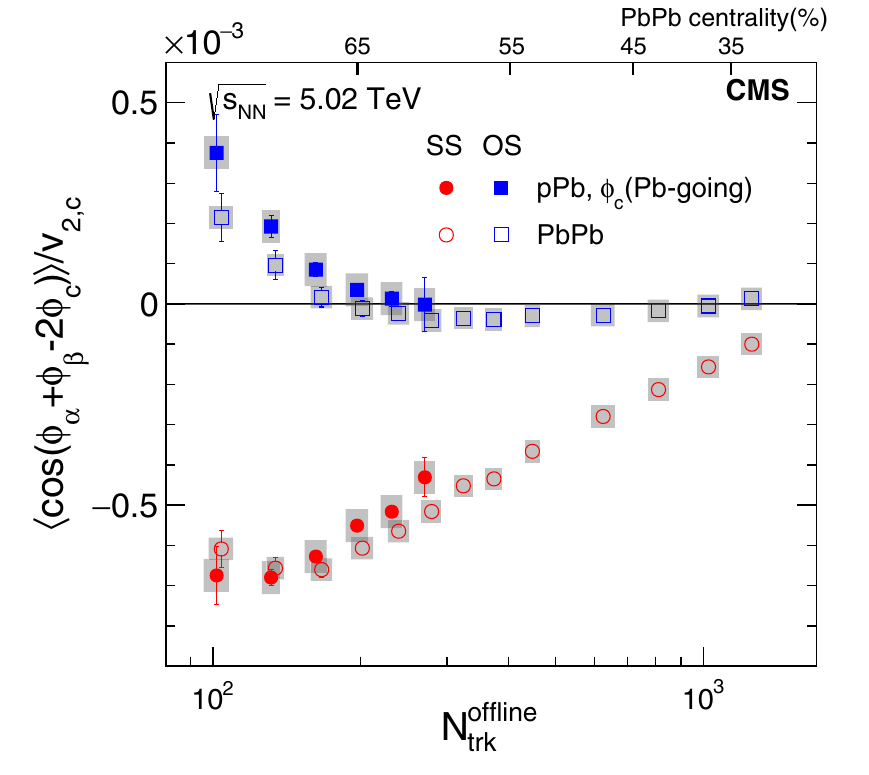} 
  \includegraphics[width=0.39\textwidth]{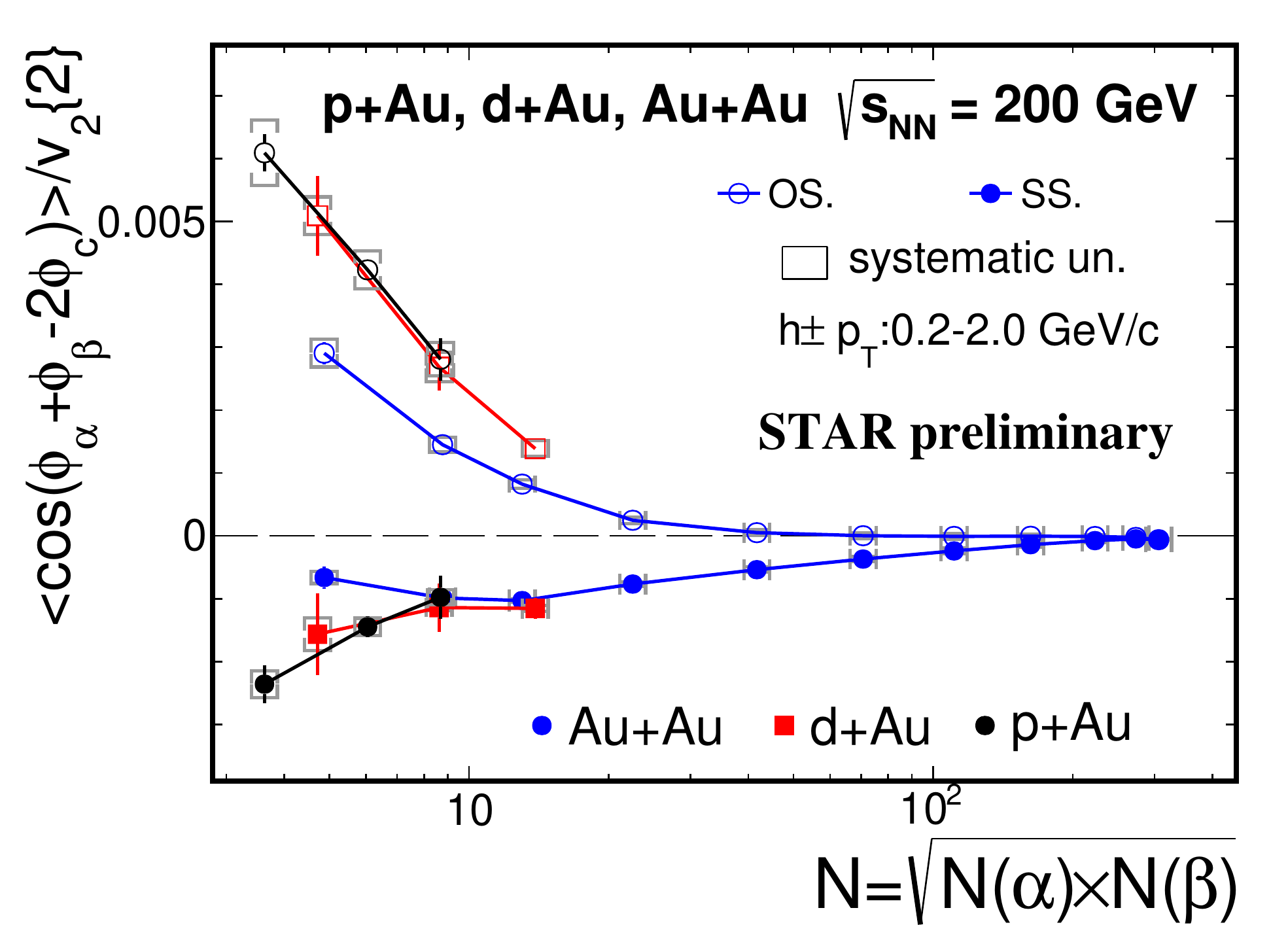} 
  \caption{(Color online) The opposite-sign (OS) and same-sign (SS) three-particle correlators in p+Pb and Pb+Pb collisions at \sNN = 5.02 TeV from CMS~\cite{Khachatryan:2016got} (upper) and in p+Au and d+Au collisions from STAR~\cite{Zhao:2017wck,Zhao:2018pnk} (lower). The CMS data are averaged over $|\eta_{\alpha}-\eta_{\beta}| < 1.6$ and plotted as a function of the offline track multiplicity, $N^{\rm offline}_{\rm trk}$. Particles $\alpha$ and $\beta$ are from the midrapidity tracker and particle $c$ from the forward/backward hadronic calorimeters for the CMS data. All three particles of the STAR data are from the TPC pseudorapidity coverage of $|\eta|<1$ with no $\eta$ gap applied; the $v_{2,c}\{2\}$ is obtained by two-particle cumulant with $\eta$ gap of $\Delta\eta > 1.0$. Statistical uncertainties are indicated by the error bars and systematic ones by the shaded regions (CMS) and caps (STAR), respectively.}   
  \label{FG_SM2}
\end{figure}

Figure~\ref{FG_SM2} upper panel shows the first $\dg$ measurements in small system p+Pb collisions at 5.02 TeV by CMS~\cite{Khachatryan:2016got}, compared with Pb+Pb at the same energy. 
Within uncertainties, the SS and OS correlators in p+Pb and Pb+Pb collisions exhibit the same magnitude and trend as a function of the event multiplicity. The CMS data further show that the $|\Delta\eta|=|\eta_{\alpha}-\eta_{\beta}|$ and multiplicity dependences of the $\dg$ correlators are similar between p+Pb and Pb+Pb collisions~\cite{Khachatryan:2016got}.
The $|\Delta\eta|$ dependence shows a traditional short-range correlation structure, a behavior also observed in the early STAR data~\cite{Abelev:2009ad}. This indicates that the correlations may come from the hadonic stage of the collisions, while the CME is expected to be a long-range correlation arising from the early stage.
The similarity seen between high-multiplicity p+Pb and peripheral Pb+Pb collisions strongly suggests a common physical origin, challenging the attribution of the observed charge-dependent correlations to the CME~\cite{Khachatryan:2016got}. 

Similar control experiments have also been performed at RHIC, using p+Au and d+Au collisions~\cite{Zhao:2017wck,Zhao:2018pnk}. Figure~\ref{FG_SM2} lower panel shows the $\gSS$ and $\gOS$ correlators as functions of particle multiplicity ($\mult$) in \pA\ and \dA\ collisions at $\snn=200$~GeV. 
Here $\mult$ is taken as the geometric mean of the multiplicities of particle $\alpha$ and $\beta$.
The corresponding \AuAu\ results are also shown for comparison. 
The trends of the correlators are similar, decreasing with increasing $\mult$. 
Similar to LHC, the small system data at RHIC are found to be comparable to Au+Au results at similar multiplicities. However, quantitative differences may exist. 
The CMS \pPb\ data are from high multiplicity collisions, overlapping with \PbPb\ data in the 30-50\% centrality range, whereas the RHIC \pdAu\ data are from minimum bias collisions, overlapping with \AuAu\ data only in peripheral centrality bins. Since the decreasing rate of $\dg$ with $\mult$ is larger in \pdAu\ than in \AuAu\ collisions, the \pdAu\ data could be quantitatively consistent with the \AuAu\ data at large $\mult$ in the range of the 30-50\% centrality. 
Given that the STAR data are preliminary and that the multiplicity coverages are different between RHIC and LHC, the similarities in the RHIC and LHC data regarding the comparisons between small-system and heavy-ion collisions are astonishing.
\section{Current status of CME measurements}
\label{sec:current}

Experimentally, there have been many efforts to reduce or eliminate backgrounds. These include:
(1) event shape selection, by varying the event-by-event $\vexe$ exploiting statistical (and dynamical) fluctuations~\cite{Adamczyk:2013kcb,Wen:2016zic},
(2) event shape engineering exploiting dynamical fluctuations in $v_2$~\cite{Schukraft:2012ah,Acharya:2017fau,Sirunyan:2017quh};
(3) comparative measurements with respect to the RP and the participant plane (PP)~\cite{Xu:2017qfs,Xu:2017zcn,ZhaoQM18} taking advantage of the geometry fluctuation effects on the PP and the magnetic field direction; and
(4) the invariant mass dependence of the $\dg$ to identify and remove the resonance decay backgrounds~\cite{Zhao:2017nfq,Zhao:2018pnk,Zhao:2017wck,ZhaoQM18,WangQM18}.
We will review these efforts in this section.

There have been several other studies related to CME that we do not cover in this review. One is to take the ratio of the measured $\dg$ to the ``expected'' elliptic flow background~\cite{Bzdak:2012ia,Adamczyk:2014mzf,Wen:2017mol}, the so-called $\kappa$ variable, and study its behavior as functions of centrality and particle species. Such a study has yielded limited insights because the expected background is not well determined. The other study is to investigate the broadness of the $\Delta S$ variable~\cite{Ajitanand:2010rc,Magdy:2017yje} and compare it to CME signal and background models. However, it is unclear whether such comparisons lead to unique conclusions~\cite{Bozek:2017plp,Feng:2018chm}.
It has been suggested~\cite{Voloshin:2010ut} that, because the Uranium (U) nucleus is strongly deformed, U+U collisions could give insights into the background issue. In very central U+U collisions, the magnetic field is negligible and the elliptic flow is appreciable because of the deformed nuclei in the initial state. This would yield appreciable $\dg$ measurement in those very central collisions. However, because the initial geometry from random orientations of the colliding nuclei is difficult to experimentally disentangle, the U+U data have so far not generated enough insights as anticipated~\cite{Wang:2012qs,Tribedy:2017hwn}.

\subsection{Event-by-event selection methods}
\label{subsec:event-by-event}

The main background sources of the $\dg$ measurements are from the $v_{2}$-induced effects. These backgrounds are expected to be proportional to $v_{2}$; see Eq.~(\ref{EQ_2}). One possible way to eliminate or suppress these $v_{2}$-induced backgrounds is to select ``spherical'' events with $\vexe=0$ exploiting the statistical and dynamical fluctuations of the event-by-event (ExE) $\vexe$. Due to finite multiplicity fluctuations, one can easily vary the shape of the final particle momentum space, which is directly related to the $v_{2}$ backgrounds~\cite{Adamczyk:2013kcb}. 

\begin{figure}[htbp!]
  \centering 
  \includegraphics[width=0.40\textwidth]{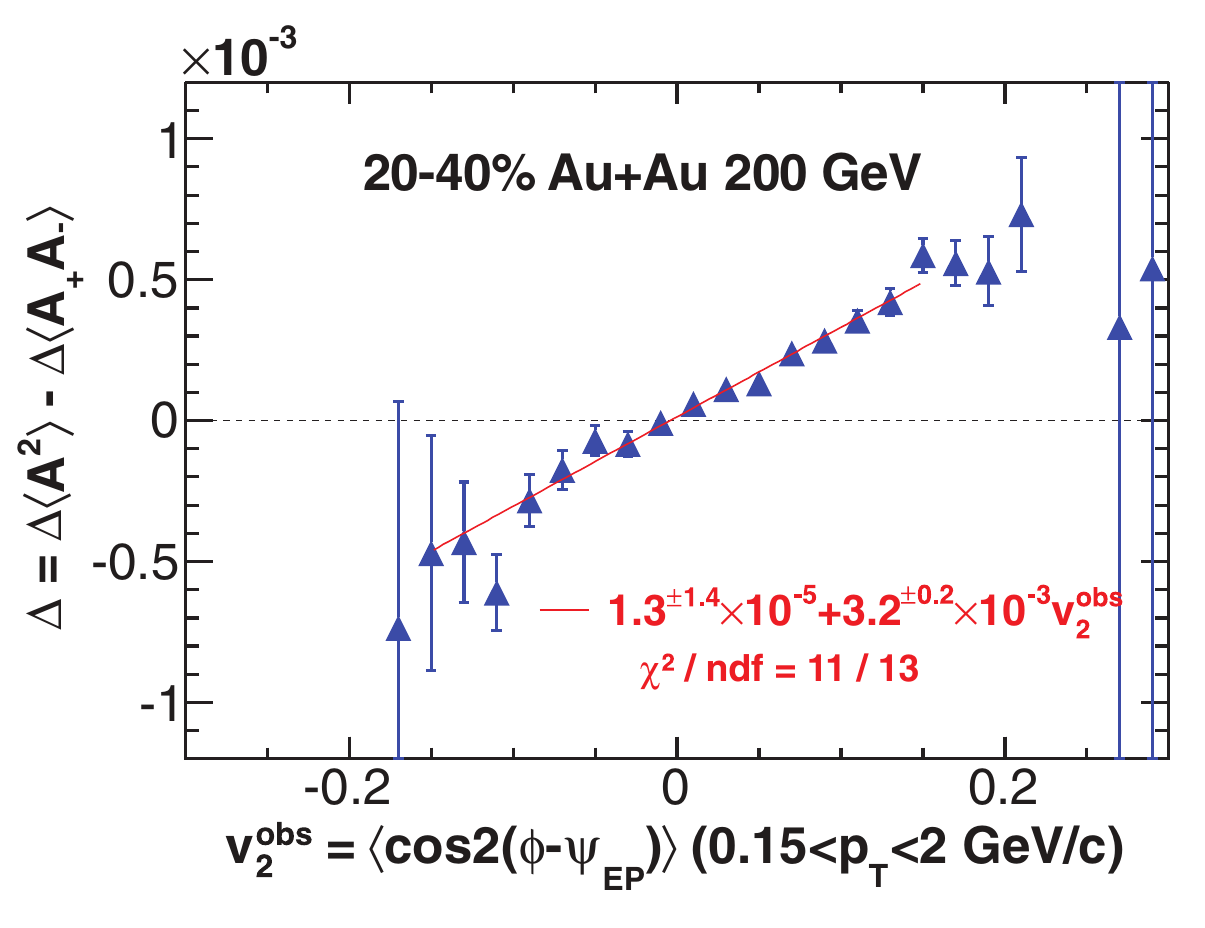} 
  \includegraphics[width=0.40\textwidth]{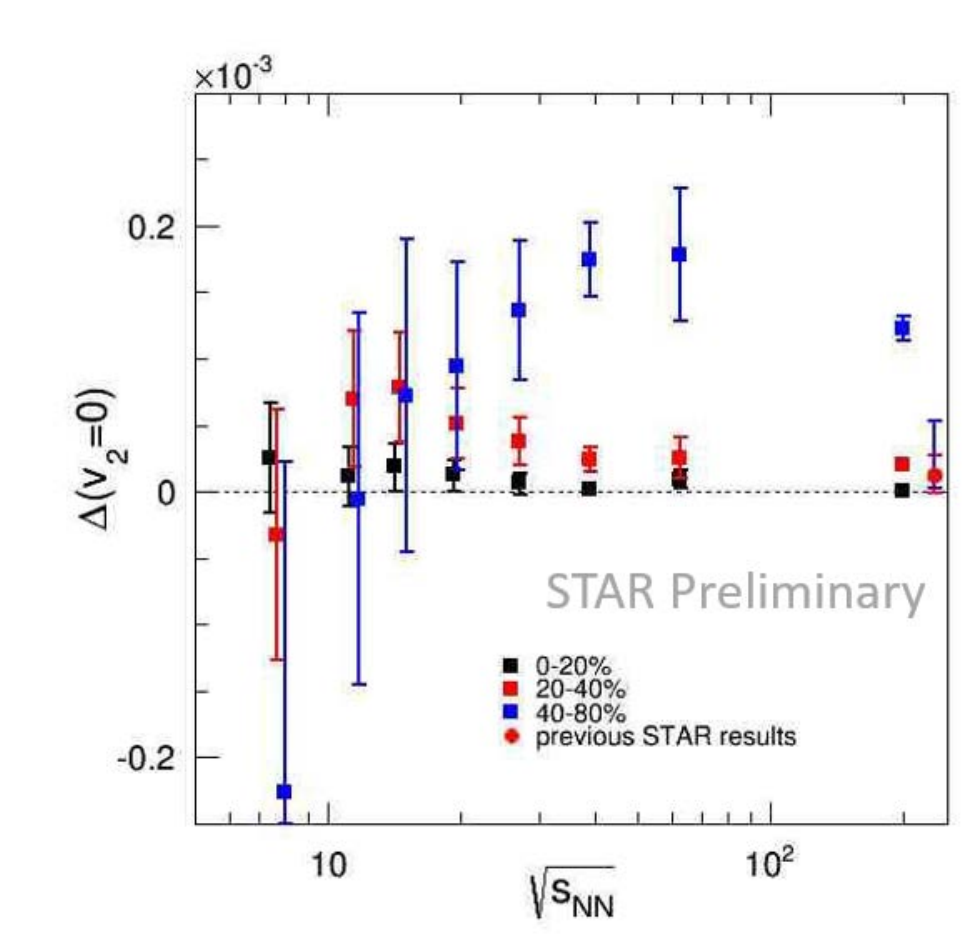} 
  \caption{(Color online) Upper: charge multiplicity asymmetry correlation ($\Delta$) as a function of $\vexe$ in 20-40\% Au+Au collisions at \sNN = 200~GeV~\cite{Adamczyk:2013kcb} from Run-4. Lower: the $\Delta$ intercept at $\vexe=0$ in various centralities of Au+Au collisions from the Beam Energy Scan data as well as from the higher statistics 200~GeV data~\cite{TuBiaoQM15,Zhao:2017ckp}.}   
  \label{FG_STARese}
\end{figure}

By using the ExE $\vexe$, STAR has carried out the first attempt to remove the backgrounds~\cite{Adamczyk:2013kcb} in their measurement of the charge multiplicity asymmetry correlations, called the $\Delta$ observable (which is similar to the $\gamma$ correlator).
The ExE $\vexe$ can be measured by the $Q$ vector method:
\begin{equation}
  \begin{split}
    &Q_n=\frac{1}{M}\sum_{j=1}^{N} w_j e^{in\phi_j}\,,\\
    &q_{n,{\rm EP}}=e^{in\psiEP}\,, \\
    &\vnexe = Q_n^{*}q_{n,{\rm EP}}\,,   \\
    &{\rm where}\;n=2,3\,.   \\
  \end{split}
  \label{EQ_ESE1}
\end{equation}
$Q_n$ sums over all particles of interest (used for the $\Delta$ variable) in each event; $\phi_{j}$ is the azimuthal angle of the $j$-th particle, and $w_{j}$ is the weight. Depending on experiments and detectors, the weights are applied in order to account for finite detector granularity or efficiency.
In Eq.~(\ref{EQ_ESE1}), $\psiEP$ is the event plane (EP) azimuthal angle, reconstructed from final-state particles, as a proxy for the PP azimuthal angle ($\psiPP$) that is not experimentally accessible.
To avoid self-correlation, particles used for the EP calculations are exclusive from the particles of interest used for $Q_2$ and $\Delta$.  
Figure~\ref{FG_STARese} upper panel shows the $\Delta$ as a function of $\vexe$ in 20-40\% Au+Au collisions at $\snn=200$~GeV~\cite{Adamczyk:2013kcb}.
A clear linear dependence is observed as expected from backgrounds.
By selecting the events with $\vexe=0$, the backgrounds in the $\Delta$ observable are largely reduced~\cite{Adamczyk:2013kcb,TuBiaoQM15,Zhao:2017ckp}. 
The intercept of a linear fit, sensitive to potential CME signals, is consistent with zero.
The lower panel of Fig.~\ref{FG_STARese} shows the extracted intercept as a function of centrality for Au+Au collisions of different beam energies~\cite{TuBiaoQM15,Zhao:2017ckp}. Positive intercepts are observed, including at beam energy of $\snn=200$~GeV with more statistics of the preliminary data.

\begin{figure}[htbp!]
  \centering 
  \includegraphics[width=0.45\textwidth]{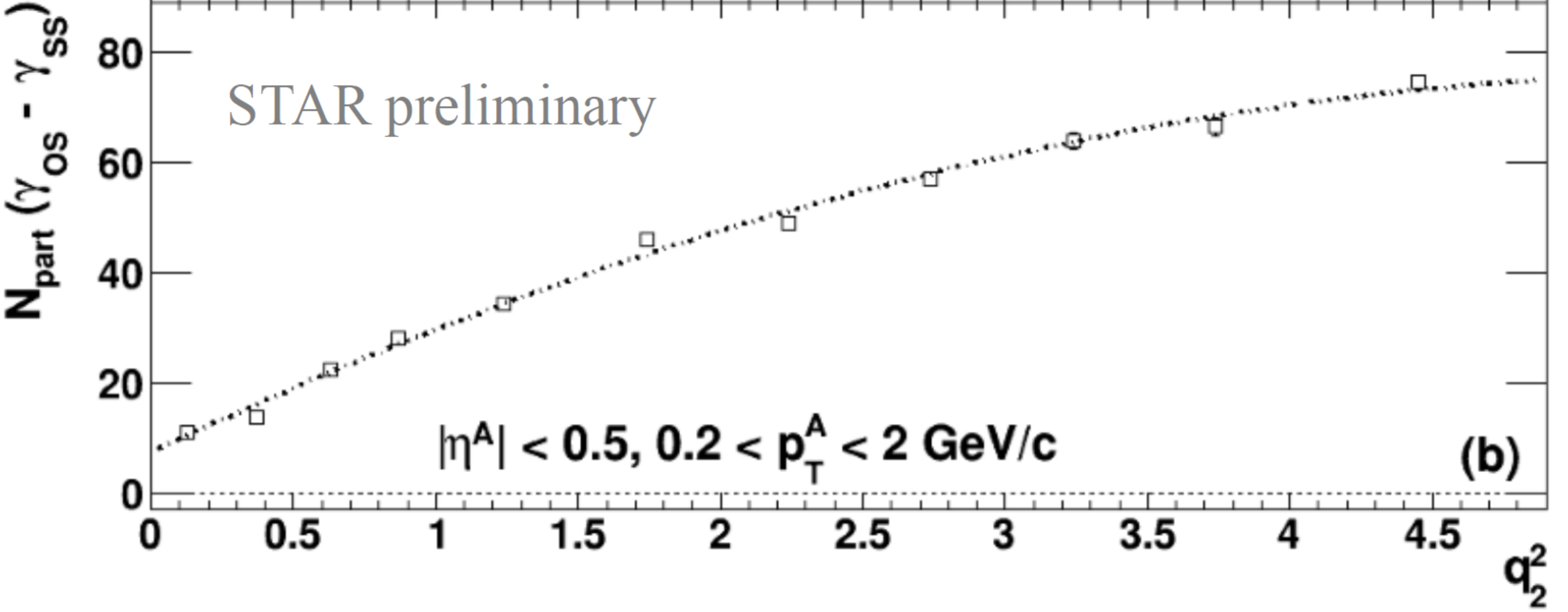} 
  \includegraphics[width=0.45\textwidth]{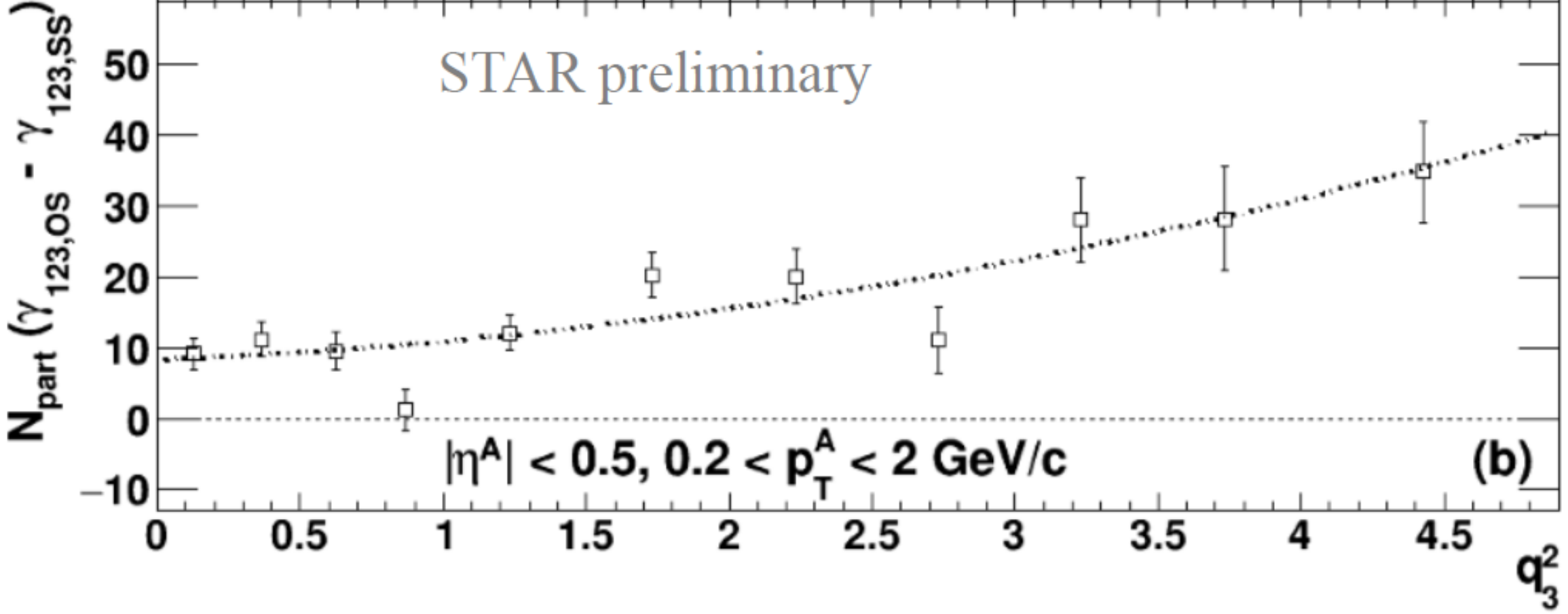} 
  \caption{(Color online) The $\dg$ correlator multiplified by the number of participants (\Npart) as a function of the ExE $q_2^2$ (upper), and that with respect to the third harmonic plane ($\dg_{123}$) as a function of $q_3^2$ in 20-60\% Au+Au collisions at \sNN = 200~GeV~\cite{WangGang_q2q3}.}   
  \label{FG_Gang}
\end{figure}

A similar method selecting events with the ExE $q_n$ variable has been proposed recently~\cite{Wen:2016zic}. Here $q_n$ is the magnitude of the second-order reduced flow vector~\cite{Adler:2002pu}, defined as:
\begin{equation}
  \begin{split}
    q_n = \sqrt{M}|Q_n|\;\;\;{\rm where}\;n=2,3\,,
  \end{split}
  \label{EQ_ESE2}
\end{equation}
and is related to $v_n$.
To suppress the $v_{2}$-induced background, a tight cut, $q_{2} = 0$, is proposed. The cut is tight because $q_{2} = 0$ corresponds to a zero $2^{nd}$ harmonic to any plane, while $\vexe = 0$ corresponds to zero $2^{nd}$ harmonic with respect to the reconstructed EP in the event.
This $q_2$ method is therefore more difficult than the ExE $v_2$ method because the extrapolation to zero $q_2$ is statistics limited and because it is unclear whether the background is linear in $q_2$ or not.
Figure~\ref{FG_Gang} shows the preliminary results from this method by STAR~\cite{WangGang_q2q3}. An extrapolation to zero $q_2$ indicates a positive intercept (see Fig.~\ref{FG_Gang} upper panel). A similar study using the third harmonic EP indicates a positive intercept as well (see Fig.~\ref{FG_Gang} lower panel), comparable in magnitude to that from the $q_2$ method.

\begin{figure}[htbp!]
  \centering 
  \includegraphics[width=0.40\textwidth]{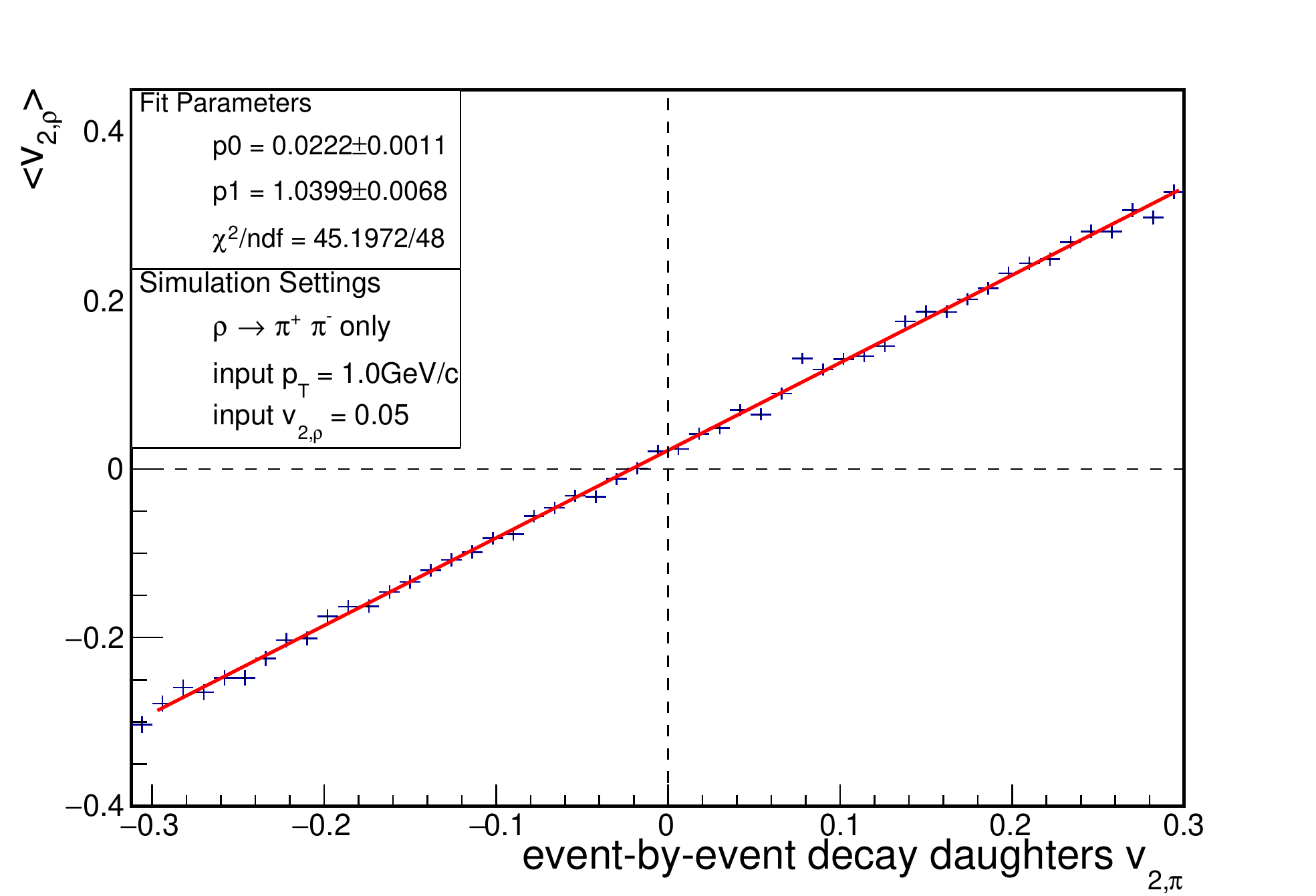}
  \includegraphics[width=0.40\textwidth]{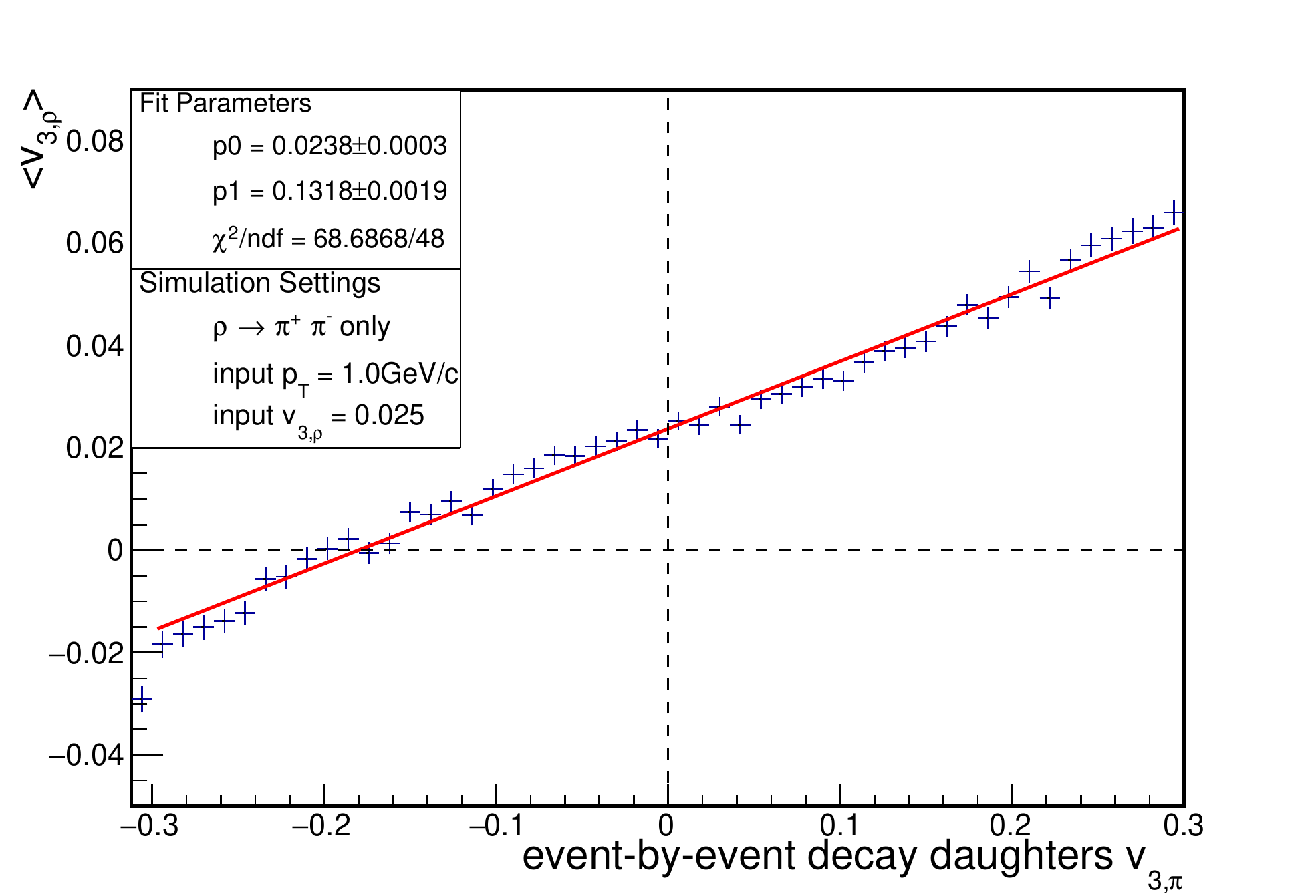}
  \caption{(Color online) $\mean{v_{2,\rho}}$ vs.~$v_{2,\pi,{\rm ebye}}$ (upper) and $\mean{v_{3,\rho}}$ vs.~$v_{3,\pi,{\rm ebye}}$ (lower) from toy-model simulations of $\rho$ resonances with fixed $p_{T,\rho}=1.0$~\gevc, $v_{2,\rho}=5$\% and $v_{3,\rho}=2.5$\%. The finite $\mean{v_{2,\rho}}$ and $\mean{v_{3,\rho}}$ values are the reasons why flow backgrounds cannot be completely removed by $v_{2,\pi,{\rm ebye}}=0$ or $v_{3,\pi,{\rm ebye}}=0$. Toy model from~\cite{Wang:2016iov}.}
  \label{FG_rhov2piv2}
\end{figure}

These methods assume the backgrounds to be linear in $v_{2}$ of the {\em final-state} particles. However, the backgrounds arise from the correlated pairs from resonance/cluster decays coupled with the $v_{2}$ of the {\em parent} sources, not that of the final-state particles. In case of resonance decays, $\dg$ depends on the $\vres$ of the resonances, not that of the decay particles or all final-state particles. Since the $v_2$ in this method is the event-by-event quantity, the resonance $\vres$ is unnecesarily zero when the final-state particle $\vexe$ is selected to be zero. This is shown in Fig.~\ref{FG_rhov2piv2} in a resonance toy model simulation~\cite{Wang:2016iov} where the average $v_n$ of the $\rho$ resonances in events with $\vnexe=0$ are found be to nonzero. It is interesting to note that the intercepts are similar for $v_2$ and $v_3$, and the slope for $v_3$ is significantly smaller than that for $v_2$. This would explain the features in Fig.~\ref{FG_Gang} where the inclusive $\dg_{123}$ is much smaller than the inclusive $\dg$ but the $q_n=0$ projection intercepts are similar. We conclude that the positive intercept results from the ExE $v_2$ and $q_2$ methods are likely still contaminated by flow backgrounds. Moreover, it is difficult, if not at all possible, to ensure the $v_{2}$ of all the background sources to be zero on event-by-event basis. Therefore, it is challenging to completely remove the flow backgrounds by using the ExE $v_{2}$ or $q_{2}$ method~\cite{Wang:2016iov}.

\subsection{Event shape engineering}
\label{subsec:ESE}

Based on the $v_2$-driven background~\cite{Wang:2009kd,Bzdak:2010fd,Schlichting:2010qia}, it is essential to explicitly investigate the $v_2$ dependence of the CME observable. One of the main diffuculties is that the conventional method of varying the $v_2$ is to select different centralities on an event-averaged basis, which will inevitably alter the initial magnetic field due to its initial-geometry dependence. However, this difficulty can be overcome by a new experimental method, called ``Event Shape Engineering" (ESE), to select events with very different $v_2$ within a narrow centrality range, where the expected CME signal is mostly independent of this event-by-event selection~\cite{Schukraft:2012ah,Acharya:2017fau,Sirunyan:2017quh}. This provides a way to decouple effects from the magnetic field and the $v_{2}$, and thus a possible solution to disentangle background contributions from potential CME signals.

In the method of ESE, instead of selecting on $\vexe$ directly, 
one uses the $Q$-vector [Eqs.~(\ref{EQ_ESE1}),~(\ref{EQ_ESE2})] to access the initial participant geometry, which
selects different event shapes from the initial-state geometry fluctuations~\cite{Schukraft:2012ah,Voloshin:2010ut,Bzdak:2011np,Acharya:2017fau, Sirunyan:2017quh}.
In particular, the ESE is performed based on the $q_{2}$ magnitude~\cite{Adler:2002pu}. This is very similar to the ExE $q_2$ method described in Sect.~\ref{subsec:event-by-event}, with one important distinction. In the ExE $q_2$ method, the $q_2$ is computed using particles of interest, whereas in ESE, the $q_2$ is computed using particles displaced away (e.g.~in pseudorapidity) from the particles of interest. Thus, the $v_2$ of the particles of interest differ for different ESE $q_2$ selections because of dynamical fluctuations of $v_2$, while the variation in $v_2$ in the ExE $q_2$ method is due to mainly statistical fluctuations.

\begin{figure}[htbp!]
  \centering 
  \includegraphics[width=6.2cm]{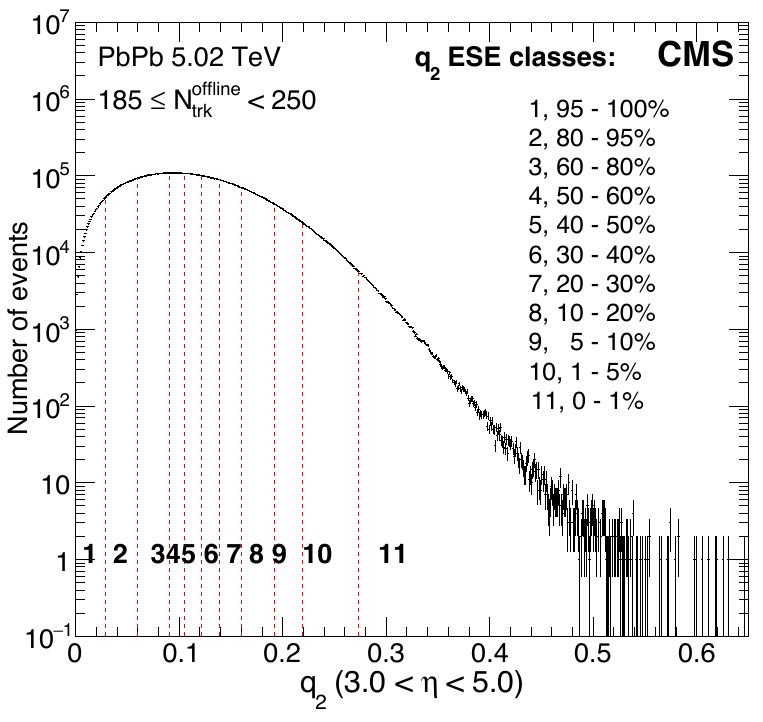} 
  \includegraphics[width=6.2cm]{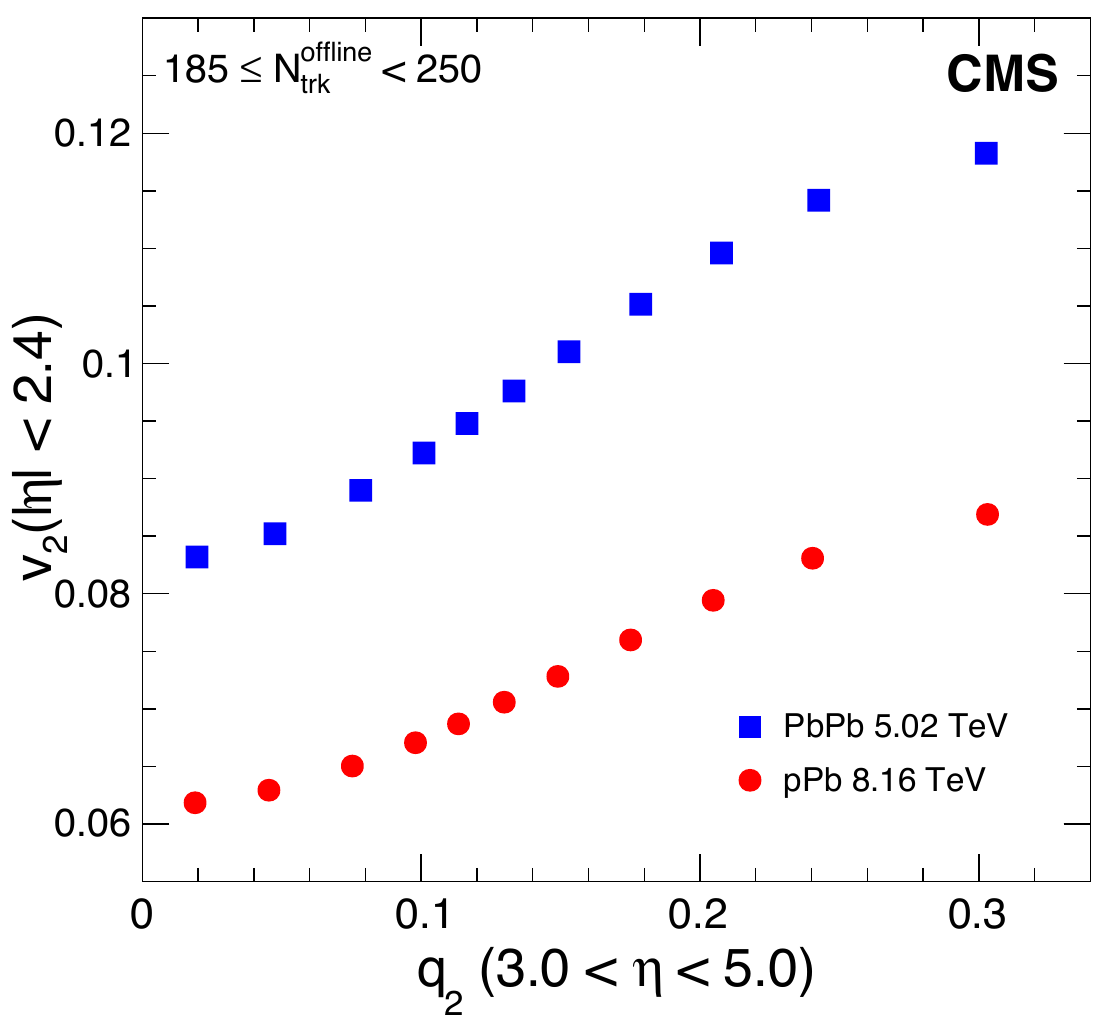}
  \caption{(Color online) Upper: the $q_{2}$ distribution in multiplicity range $185 \leq N^{\rm offline}_{\rm trk} < 250$ in Pb+Pb collisions. Red dashed lines represent the selection used to divide the events into multiple $q_{2}$ classes. Lower: the correlation between $v_{2}$ and $q_{2}$ in p+Pb and Pb+Pb collisions based on the $q_{2}$ selections of the events~\cite{Sirunyan:2017quh}.}
  \label{FG_CMSeseA}
\end{figure}

Figure~\ref{FG_CMSeseA} (upper) shows the $q_{2}$ distribution in Pb+Pb collisions from the CMS Collaboration~\cite{Sirunyan:2017quh}. Events within a narrow multiplicity range are divided into several classes with each corresponding to a fraction of the full distribution, 
where the 0-1\% represents the class with the largest $q_{2}$ value. In Fig.~\ref{FG_CMSeseA} (lower), the average $v_2$ values at mid-rapidity are presented in each selected $q_2$ class, where the strong proportionality between these two quantities suggests their underlying correlation from the initial-state geometry~~\cite{Sirunyan:2017quh}. Therefore, the $\dg$ correlator can be studied as a function of $v_2$ explicitly using the $q_2$ selections. 

\begin{figure}[htbp!]
  \includegraphics[width=0.4\textwidth]{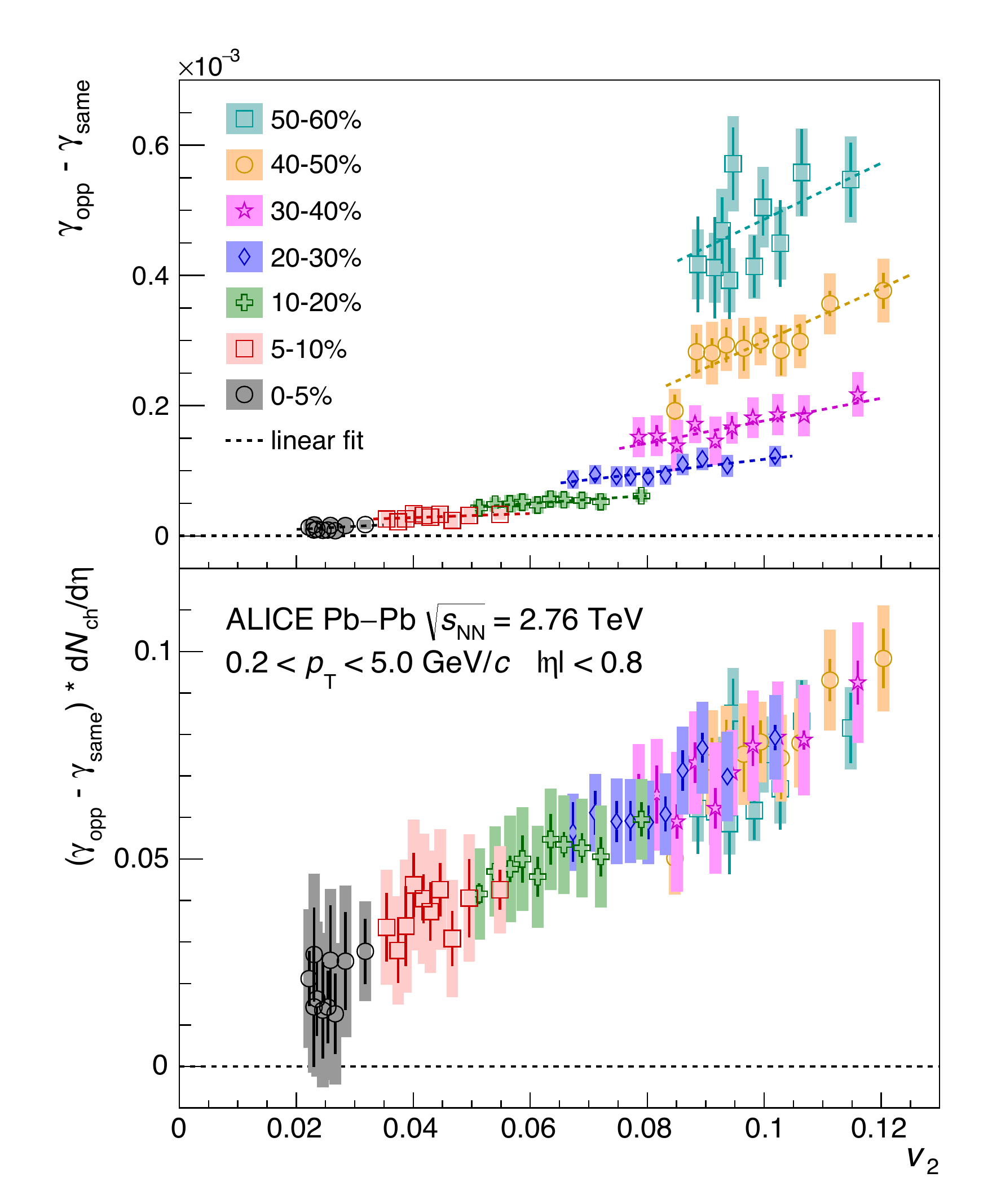} 
  \caption{(Color online) The $\dg$ correlator (upper) and the charged-particle density scaled correlator $\dg \cdot dN_{ch}/d\eta$ (lower) as functions of $v_{2}$ for shape-selected events by $q_{2}$ for various centrality classes in Pb+Pb collisions by ALICE~\cite{Acharya:2017fau}. Error bars (shaded boxes) represent the statistical (systematic) uncertainties.}
  \label{FG_ALICEeseB_1}
\end{figure}

The $\dg$ correlator has been studied as a function of $v_2$ using the ESE method in different centrality classes in Pb+Pb collisions from the ALICE Collaboration~\cite{Acharya:2017fau}, shown in Fig.~\ref{FG_ALICEeseB_1} (upper). In order to remove the trivial multiplicity dilution effect, the correlator $\dg$ that is scaled by the charge-particle density ($dN_{ch}/d\eta$) in a given centrality range, is also shown in Fig.~\ref{FG_ALICEeseB_1} (lower). The data indicate a strong linear dependence on the measured $v_2$, where different centralities fall onto the same linear trend after the multiplicity scaling. This observation is qualitatively consistent with a background scenario, i.e., local charge conservation coupled with anisotropic flow~\cite{Wang:2009kd,Pratt:2010zn,Bzdak:2010fd,Hori:2012kp,Wang:2016iov}; see Eq.~(\ref{EQ_2}).

\begin{figure}[htbp!]
  \includegraphics[width=0.42\textwidth]{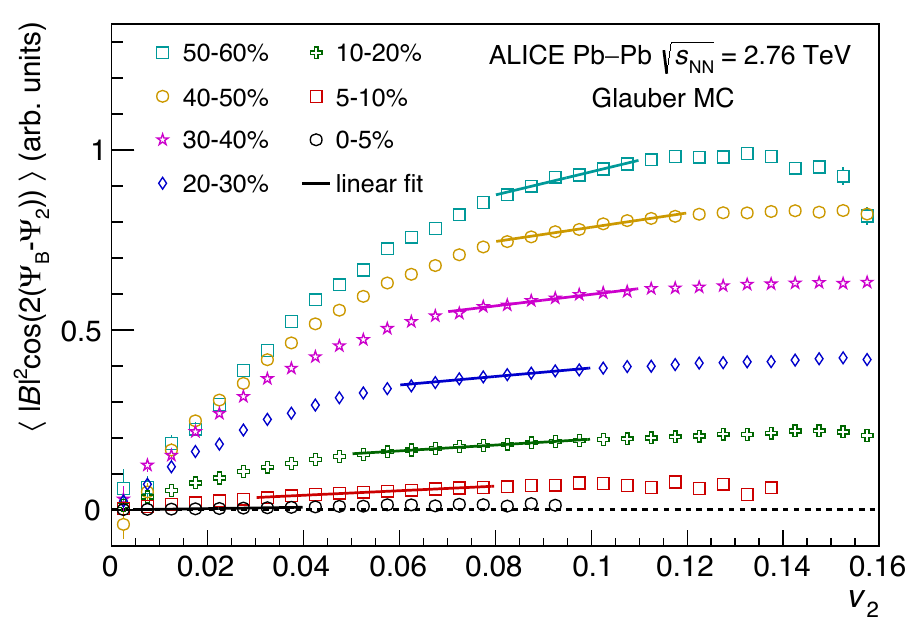}
  \caption{(Color online) The expected dependence of the CME signal on $v_{2}$ for various centrality classes from a MC-Glauber simulation~\cite{Miller:2007ri}. The solid lines depict linear fits based on the $v_{2}$ variation observed within each centrality interval~\cite{Acharya:2017fau}.}
  \label{FG_ALICEeseB_2}
\end{figure}

As argued earlier, the advantage of using the ESE is to independently evaluate the $v_2$-dependent background from the $\dg$ correlator without significantly changing the CME signal due to the magnetic field. However, this assumption is not exactly true as the observable signal of the CME also depends on how precise the $v_2$ can be measured; in other words, the signal extraction depends on the $v_2$ resolution. From the study of the ALICE experiment~\cite{Acharya:2017fau}, the signal dependence on the $v_2$ (resolution) has been explicitly investigated using different {\em Monte Carlo} (MC) Glauber calculations, shown in Fig.~\ref{FG_ALICEeseB_2}. Specifically, the CME signal is assumed to be proportional to $\mean{|\textbf B|^{2} \cos2(\psiB - \psi_{2})}$, where $|\textbf B|$ and $\psiB$ are the magnitude and azimuthal direction of the magnetic field. As one can see, the dependence is stronger in small $v_2$ region than in large $v_2$, and in most central or most peripheral events than in mid-central. Therefore, with the input of the signal dependence on $v_2$, the residual CME signal can be extracted based on the different dependences of signal and background correlation on the measured $v_2$. 

To extract the contribution of the possible CME signal from the current $\dg$ measurements, a linear function is fit to the data:
\begin{equation}
  \begin{split}
    F_{1}(v_{2}) = p_{0}(1+p_{1}(v_{2}-\mean{v_{2}})/\mean{v_{2}})\,. 
  \end{split}
  \label{EQ_ESE3}
\end{equation}
Here $p_{0}$ accounts for an overall scale, and the $p_{1}$ is the normalized slope, reflecting the $v_{2}$ dependence. In a pure background scenario, the $\dg$ correlator is proportional to $v_{2}$ and the $p_{1}$ parameter is expected to be unity, thus Eq.~\ref{EQ_ESE3} is reduced to $F_{1}(v_{2}) = p_{0}v_{2}/\mean{v_{2}} \propto v_{2} $. 
On the other hand, a significant CME contribution would result in a non-zero intercept at $v_{2}=0$ of the linear functional fits. 

In a two-component model with signal and background, a measured observable ($O_{m}$) can be expressed as: 
\begin{equation}
  \begin{split}
    \frac{S}{S+B}\times O_{S} + \frac{B}{S+B}\times O_{B} = O_{m}\,, \\  
  \end{split}
  \label{EQ_ESE4}
\end{equation}
$O_{S}$ and $O_{B}$ are the values of the observable $O$ from signal and background respectively, and $\frac{S}{S+B}$ represents the fraction of signal contribution in the measurement.
The $p_{1}$ from the fit to the measured data is thus a combination of CME signal slope ($p_{1,\rm{sig}} = p_{1,\rm{MC}}$) and the background slope ($ p_{1,\rm{bkg}} \equiv 1$): 
\begin{equation}
  \begin{split}
    &f_{\rm{CME}}\times p_{1,\rm{sig}} + (1-f_{\rm{CME}})\times p_{1,\rm{bkg}} = p_{1,\rm{data}}\,,
  \end{split}
  \label{EQ_ESE5}
\end{equation}
where $f_{\rm{CME}} = \frac{\dg_{\rm{CME}}}{\dg_{\rm{CME}}+\dg_{\rm{bkg}}}$ represents the CME fraction to the $\dg$ correlator from the measurements, and $p_{1,\rm{MC}}$ is the slope parameter from the MC calculations in Fig.~\ref{FG_ALICEeseB_2}.

\begin{figure}[htbp!]
  \includegraphics[width=0.4\textwidth]{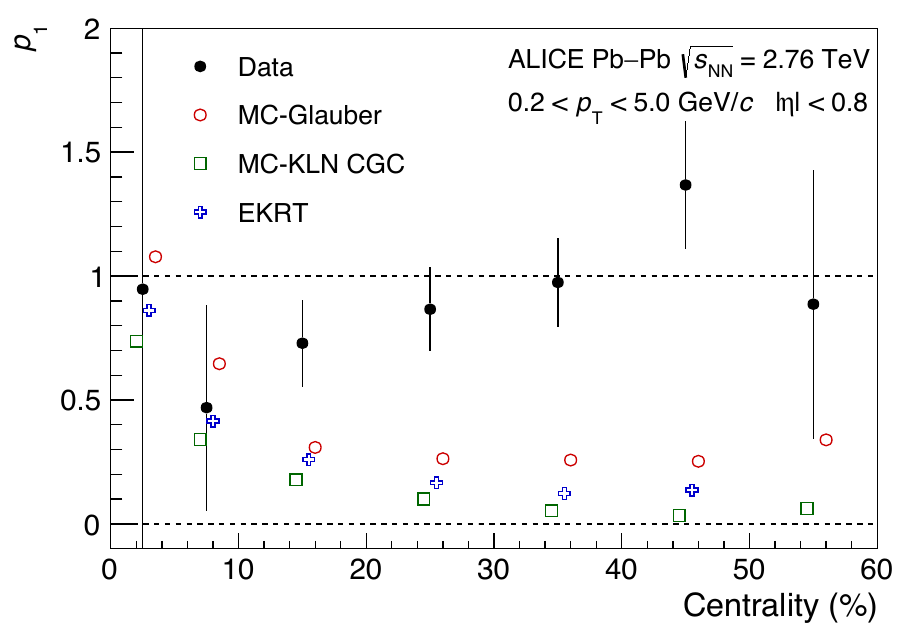} 
  \includegraphics[width=0.4\textwidth]{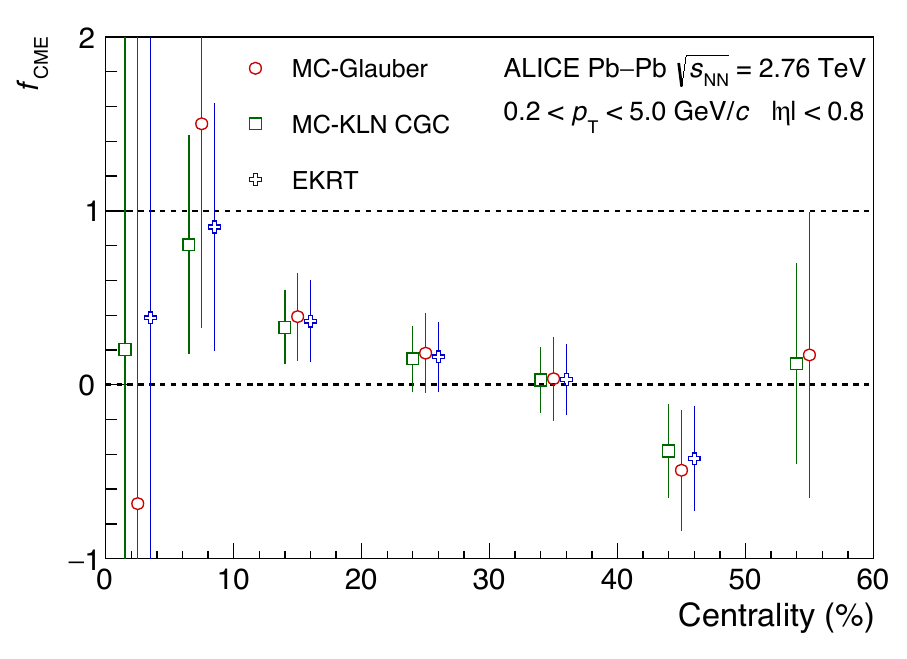}
  \caption{(Color online) Upper: centrality dependence of the $p_{1}$ parameter from a linear fit to the $\dg$ correlator in Pb+Pb collisions from ALICE and from linear fits to the CME signal expectations from MC-Glauber~\cite{Miller:2007ri}, MC-KLN CGC~\cite{Drescher:2007ax,ALbacete:2010ad}, and EKRT~\cite{Niemi:2015qia} models. Lower: centrality dependence of the CME fraction extracted from the slope parameter of fits to data and different models. Points from MC simulations are slightly shifted along the horizontal axis for better visibility. Only statistical uncertainties are shown. From Ref.~\cite{Acharya:2017fau}.}
  \label{FG_ALICEeseC}
\end{figure}

Figure~\ref{FG_ALICEeseC} (upper) shows the centrality dependence of $p_{1,\rm{data}}$ from data and $p_{1,\rm{MC}}$ from signal expectations based on MC-Glauber, MC-KLN CGC and EKRT models~\cite{Acharya:2017fau}. Figure~\ref{FG_ALICEeseC} (lower) presents the estimate $f_{\rm{CME}}$ from the three models. The $f_{\rm{CME}}$ extracted from central (0-10\%) and peripheral (50-60\%) events have large statistical uncertainties. Combining the data from 10-50\% centrality with an assumption of a constant CME contribution, it gives a value of $f_{\rm{CME}} = 0.10 \pm 0.13$, $0.08 \pm 0.10$, and $0.08 \pm 0.11$ 
for the MC-Glauber, MC-KLN CGC and EKRT models, respectively. 
These results are consistent with zero CME fraction within the uncertainty, and correspond to upper limits on $f_{\rm{CME}}$ of 33\%, 26\% and 29\%, respectively, at 95\% confidence level (CL) for the centrality range of 10-50\%~\cite{Acharya:2017fau}. 

\begin{figure}[htbp!]
  \includegraphics[width=0.4\textwidth]{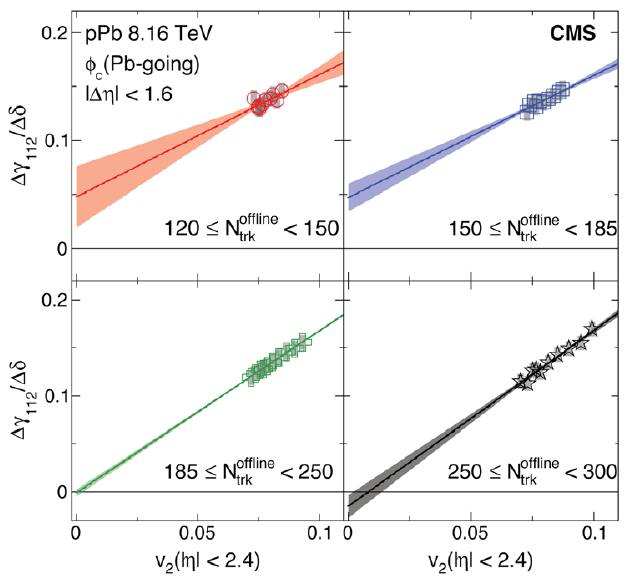} 
  \includegraphics[width=0.4\textwidth]{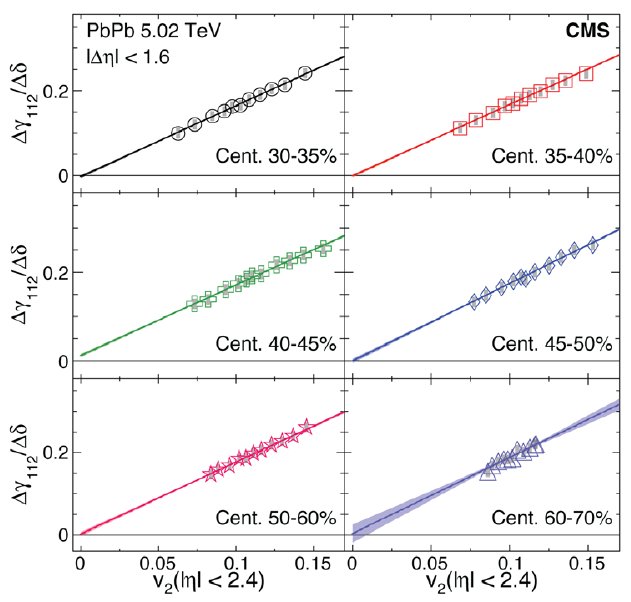} 
  \caption{(Color online) The ratio between $\dg$ ($\dg_{112}$) and $\Delta\delta$ correlators, $\dg/\Delta\delta$, averaged over $|\Delta\eta|< 1.6$ as a function of $v_{2}$ evaluated in each $q_{2}$ class, for different multiplicity and centrality ranges in p+Pb (upper) and Pb+Pb (lower) collisions~\cite{Sirunyan:2017quh}.}
  \label{FG_CMSESEA}
\end{figure}
\begin{figure}[htbp!]
  \includegraphics[width=0.42\textwidth]{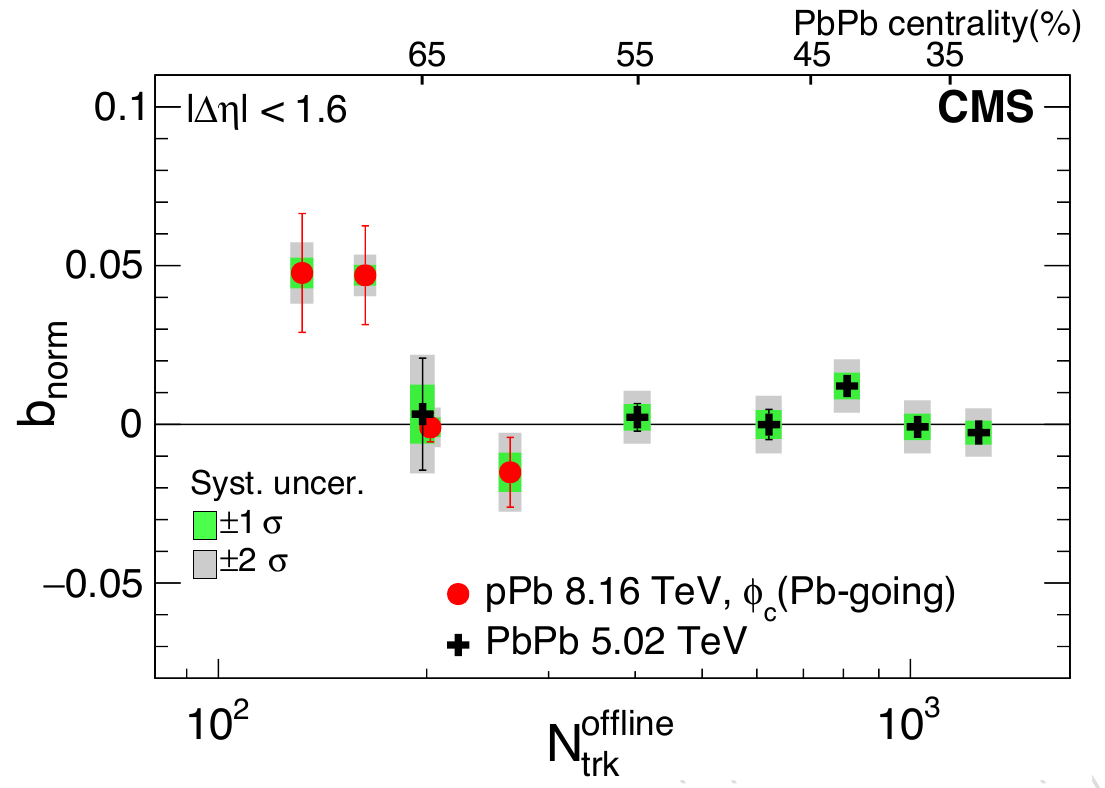} 
  \includegraphics[width=0.40\textwidth]{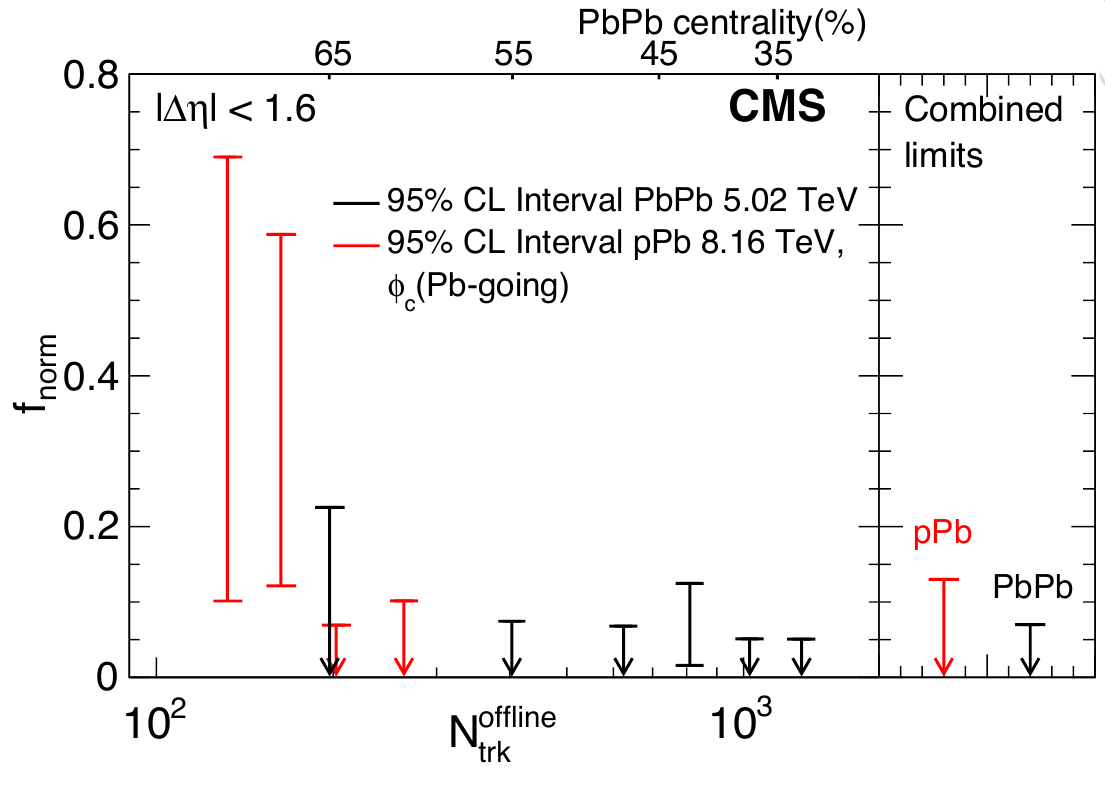} 
  \caption{(Color online) Extracted intercept parameter $b_{\rm{norm}}$ (upper) and their corresponding upper limits of the fraction of the $v_{2}$-independent $\dg$ correlator component (lower), averaged over $|\Delta\eta|< 1.6$, as a function of $N^{\rm offline}_{\rm trk}$ in p+Pb and Pb+Pb collisions from the CMS Collaboration~\cite{Sirunyan:2017quh}.}
  \label{FG_CMSESEB}
\end{figure}

The above analysis method is model-dependent, which relies on precise modeling of the correlation between magnetic field and $v_2$ in a given centrality range. Another approach, adopted by CMS, is to select very narrow centrality ranges with wide $v_2$ coverage~\cite{Sirunyan:2017quh}. The signal and background contribution to the $\gamma$ correlator can be separated as~\cite{Bzdak:2012ia}: 
\begin{equation}
  \begin{split} 
    \gamma & =\kappa_{2} \delta v_{2} + \gamma_{\rm{CME}}\,,   \\
    \delta &\equiv \mean{\cos(\phi_{\alpha}-\phi_{\beta})}\,. \\
  \end{split}
  \label{EQ_ESE6}
\end{equation}
Here, $\delta$ represents the charge-dependent two-particle azimuthal correlator and $\kappa_{2}$ is a parameter independent of $v_{2}$, mainly determined by the kinematics and acceptance of particle detection~\cite{Bzdak:2012ia}. Using the ESE to select events with different $v_{2}$, the above Eq.~(\ref{EQ_ESE6}) can be explicitly tested and the $v_2$-independent component of the $\gamma$ correlator ($\gamma_{\rm{CME}}$), which is related to the CME signal, can be extracted. 
The charge-independent background sources are eliminated by taking the difference of the correlators ($\gamma, \delta$) between same- and opposite-sign pairs, as was done in Ref.~\cite{Acharya:2017fau}. Therefore, Eq.~(\ref{EQ_ESE6}) becomes:
\begin{equation}
  \begin{split}
    \dg =\kappa_{2} \Delta\delta v_{2} + \Delta\gamma_{\rm{CME}}\,.  \\ 
  \end{split}
  \label{EQ_ESE7}
\end{equation}

From the ESE, it is assumed that the $\Delta\delta$ correlator is independent of $v_2$, while it has been found that it is not the case for peripheral events, mainly due to the multiplicity bias from the $q_2$ selection~\cite{Sirunyan:2017quh}. Therefore, in order to remove the $v_2$ dependence on $\Delta\delta$ correlator, both sides of Eq.~(\ref{EQ_ESE7}) are divided by $\Delta\delta$ and the equation can be simplified into  
\begin{equation}
  \begin{split}
    \dg/\Delta\delta = a_{\rm{norm}} v_{2} + b_{\rm{norm}}\,,  \\ 
  \end{split}
  \label{EQ_ESE8}
\end{equation}
where $b_{\rm{norm}}$ represents the $v_2$-independent component (scaled by $\Delta\delta$) that could be caused by the contribution of a CME signal.

Figure~\ref{FG_CMSESEA} shows the ratio of $\Delta\gamma/\Delta\delta$ as function of $v_{2}$ for different multiplicity ranges
in p+Pb (upper) and for different centrality ranges in Pb+Pb (lower) collisions~\cite{Sirunyan:2017quh} with linear fits and their statistical uncertainty bands. The extracted values of the intercept parameter $b_{\rm{norm}}$ are shown as a function of event multiplicity in Fig.~\ref{FG_CMSESEB} (upper).
Within statistical and systematic uncertainties, no significant positive value of $b_{\rm{norm}}$ is observed. Result shows that the $v_{2}$-independent contribution to the $\dg$ correlator is consistent with zero, which suggests the underlying mechanism of the observed charge-dependent correlation is due to a background-only scenario~\cite{Sirunyan:2017quh}.
Based on the assumption of a nonnegative CME signal, the upper limit of the $v_{2}$-independent fraction in the $\dg$ correlator is obtained from the Feldman-Cousins approach~\cite{Feldman:1997qc} with the measured statistical and systematic uncertainties. 
Figure~\ref{FG_CMSESEB} (lower) shows the upper limit of the fraction $f_{\rm{norm}}$, 
the ratio of the $b_{\rm{norm}}$ value to the value of $\mean{\dg}/\mean{\Delta\delta}$, as a function of event multiplicity at 95\% CL. 
The fraction of the $v_{2}$-independent component of the $\dg$ correlator is less than 8-15\% for most of the multiplicity or centrality ranges. 
The combined limits from all presented multiplicities and centralities are also shown in p+Pb and Pb+Pb collisions in Fig.~\ref{FG_CMSESEB} (lower).
An upper limit on the $v_{2}$-independent fraction of the $\dg$ correlator, or possibly the CME signal contribution, 
is estimated to be 13\% in p+Pb and 7\% in Pb+Pb collisions, at 95\% CL. 
The results are consistent with a $v_{2}$-dependent background-only scenario, 
posing a significant challenge to the search for the CME in heavy ion collisions using three-particle azimuthal correlations~\cite{Sirunyan:2017quh}.


\subsection{Measurements with respect to RP and PP}
\label{subsec:reactionplane}

The CME-induced charge separation is driven by the magnetic field, and is therefore the strongest along the magnetic field direction. The major background to the CME is related to the elliptic flow anisotropy, determined by the participant geometry, and is therefore the largest with respect to the $\psiPP$. The magnetic field direction and the PP direction are different. These facts led to the novel idea to determine the CME signal (and flow background) from $\dg$ measurements with respect to the RP and PP in the same collision event~\cite{Xu:2017qfs}.

\begin{figure}[htbp!]
  \centering 
  \includegraphics[width=7.2cm]{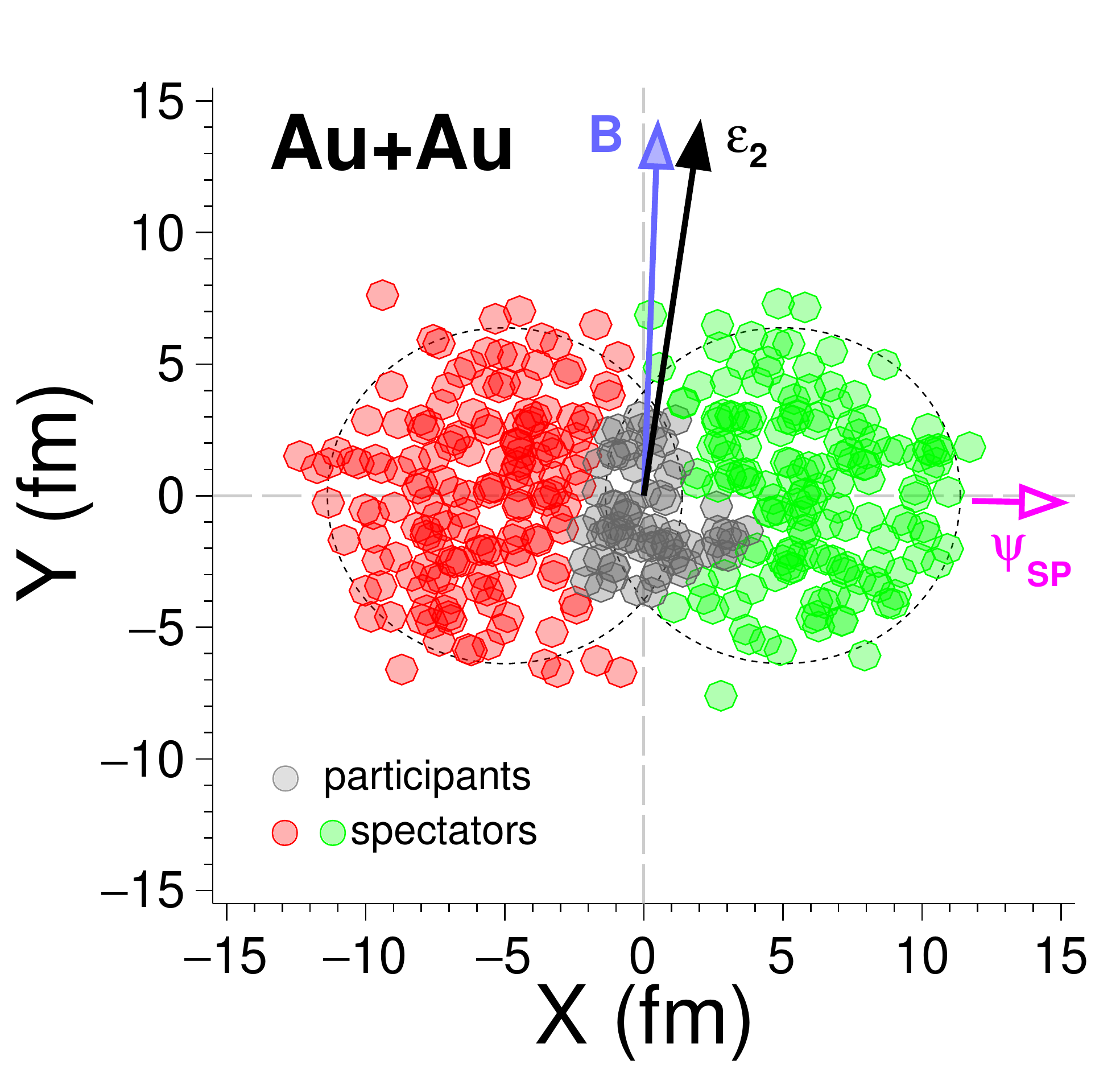} 
  \caption{(Color online) Single-event display from a MC Glauber simulation of a mid-central Au+Au collision at 200 GeV. The gray markers indicate participating nucleons, and the red (green) markers indicate the spectator nucleons traveling in positive (negative) $z$ direction. The blue arrow indicates the magnetic field direction. The long axis of the participant zone (eccentricity) is shown as the black arrow. The magenta arrow shows the direction determined by the spectator nucleons.}
  \label{FG_RP1}
\end{figure}

In general, the $\psiB$ and $\psiPP$ are correlated with the $\psiRP$, and therefore are indirectly correlated with each other. 
While the magnetic field is mainly produced by spectator protons, their positions fluctuate, so $\psi_{B}$ is not always perpendicular to the $\psiRP$. 
The position fluctuations of participant nucleons and spectator protons are independent, thus $\psiPP$ and $\psi_{B}$ fluctuate independently about $\psiRP$. Figure~\ref{FG_RP1} depicts the various azimuthal directions in the overlap transverse plane from a single MC Glauber event in mid-central Au+Au collision at 200 GeV.

\begin{figure*}[htbp!]
  \centering 
  \includegraphics[width=0.8\textwidth]{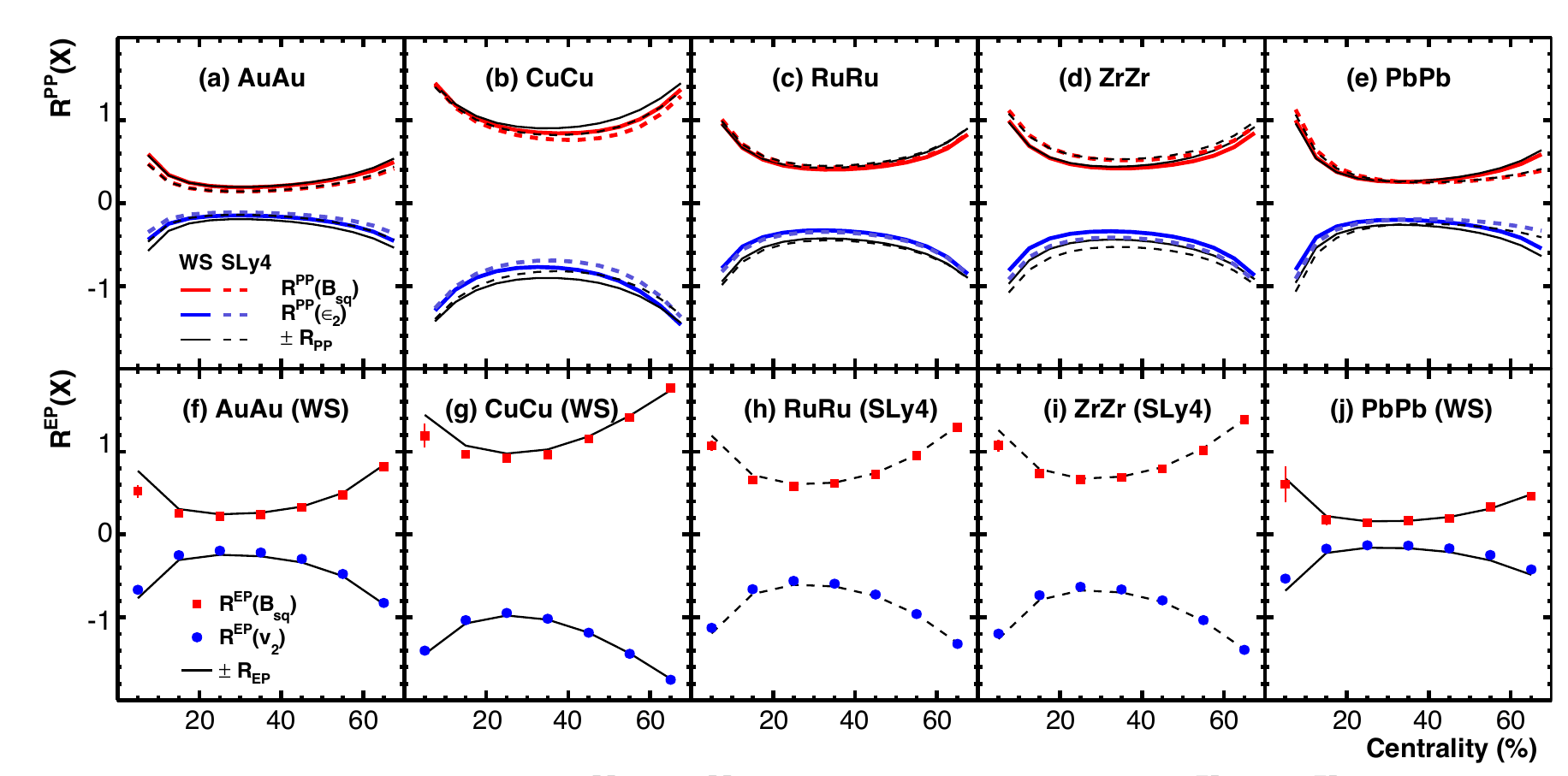} 
  \caption{(Color online) Relative differences $R(\epsilon_{2})$, $R(\Bsq)$ from MC Glauber model (upper) and $R(v_{2})$, $R(\Bsq)$ from AMPT (lower) for (a,f) Au+Au, (b,g) Cu+Cu, (c,h) Ru+Ru, and (d,i) Zr+Zr at RHIC, and (e,j) Pb+Pb at the LHC~\cite{Xu:2017qfs}. Both the Woods-Saxon and DFT-calculated~\cite{Xu:2017zcn} densities are shown for the MC Glauber calculations, while the used density profiles are noted for the AMPT results.}
  \label{FG_RP2}
\end{figure*}

The eccentricity of the transverse overlap geometry is related to the PP. It yields the largest $v_2\{{\rm PP}\}$. The $v_2$ with respect to the RP is smaller, by the factor of $a\equiv\langle\cos2(\psiPP-\psiRP)\rangle$ given by the relative angle between RP and PP. Because of fluctuations~\cite{Alver:2006wh}, the PP and RP do not coincide, so $a$ has a value always smaller than unity.
The magnetic field effect for CME, $\Bsq\{\psi\} \equiv \mean{(eB/m_{\pi}^{2})^{2}\cos2(\psi_{B}-\psi)}$, is, on the other hand, strongest along the RP direction because the magnetic field is mainly generated by the spectator protons. The effect is smaller along the PP, again by the same factor $a$.
The relative difference 
\begin{equation}
  R(X)\equiv2\cdot\frac{X\{\psiRP\}-X\{\psiPP\}}{X\{\psiRP\}-X\{\psiPP\}}
  \label{EQ_RP3}
\end{equation}
in the eccentricity (i.e.~$X$ is $\epsilon_{2}$) and magnetic field strength (i.e.~$X$ is $\Bsq$) are the opposite. Namely
\begin{equation}
  R(\Bsq)=-R(\epsilon_{2})=2(1-a)/(1+a)\,. 
  \label{EQ_RP4}
\end{equation}
This is verified by MC Glauber model calculations~\cite{Xu:2014ada,Zhu:2016puf} for various collision systems, shown in the upper panels of Fig.~\ref{FG_RP2}~\cite{Xu:2017qfs}.
The AMPT~\cite{Lin:2004en,Lin:2001zk} simulations using the reconstructed EP, shown in the lower panels of Fig.~\ref{FG_RP2}, also confirm the conclusion~\cite{Xu:2017qfs}.

The $\psiRP$, $\psiPP$ and $\epsilon_{2}$ are all theoretical concepts, and cannot be experimentally measured. Usually 1st-order harmonic EP from zero-degree calorimeters (ZDC), which measure spectator neutrons~\cite{Reisdorf:1997flow,Abelev:2013cva,Adamczyk:2016eux}, is a good proxy for $\psiRP$. As a proxy for $\psiPP$, the 2nd-order harmonic EP ($\psiEP$) reconstructed from final-state particles is used. Since $v_{2}$ is generally proportional to $\epsilon_{2}$, one can obtain the factor $a$ by
\begin{equation}
  a=v_{2}\{\psiRP\}/v_{2}\{\psiEP\}\,.
  \label{EQ_RP5}
\end{equation}

The $\dg$ variable contains CME signal and the $v_{2}$-induced background:
\begin{equation}
  \begin{split}
    \dg\{\psi\} = {\rm CME}(\Bsq\{\psi\}) + {\rm BKG}(v_{2}\{\psi\})\,.	
  \end{split}
  \label{EQ_RP7}
\end{equation}
Assuming the ${\rm CME}(\Bsq\{\psi\})$ is proportional to $\Bsq$ and ${\rm BKG}(v_{2}\{\psi\})$ is proportional to $v_{2}$, one obtains the relative CME signal to background contribution:
\begin{equation}
  r \equiv \frac{{\rm CME}(\Bsq\{\psiRP\})}{{\rm BKG}(v_{2}\{\psiEP\})}
  \approx \frac{R(v_{2})-R(\dg)}{R(v_{2})+R(\dg)}\,. 
  \label{EQ_RP9}
\end{equation}
where the $R(X)$ definition is given by Eq.~(\ref{EQ_RP3}).
The CME signal fraction in the measurements with respect to $\psiEP$ is
\begin{equation}
  f_{\rm CME}^{\rm EP} = {\rm CME}(\Bsq\{\psiEP\})/\dg\{\psiEP\}=r/(r+1/a)\,.
  \label{EQ_RP10}
\end{equation}

\begin{figure}
  \centering 
  \includegraphics[width=0.4\textwidth]{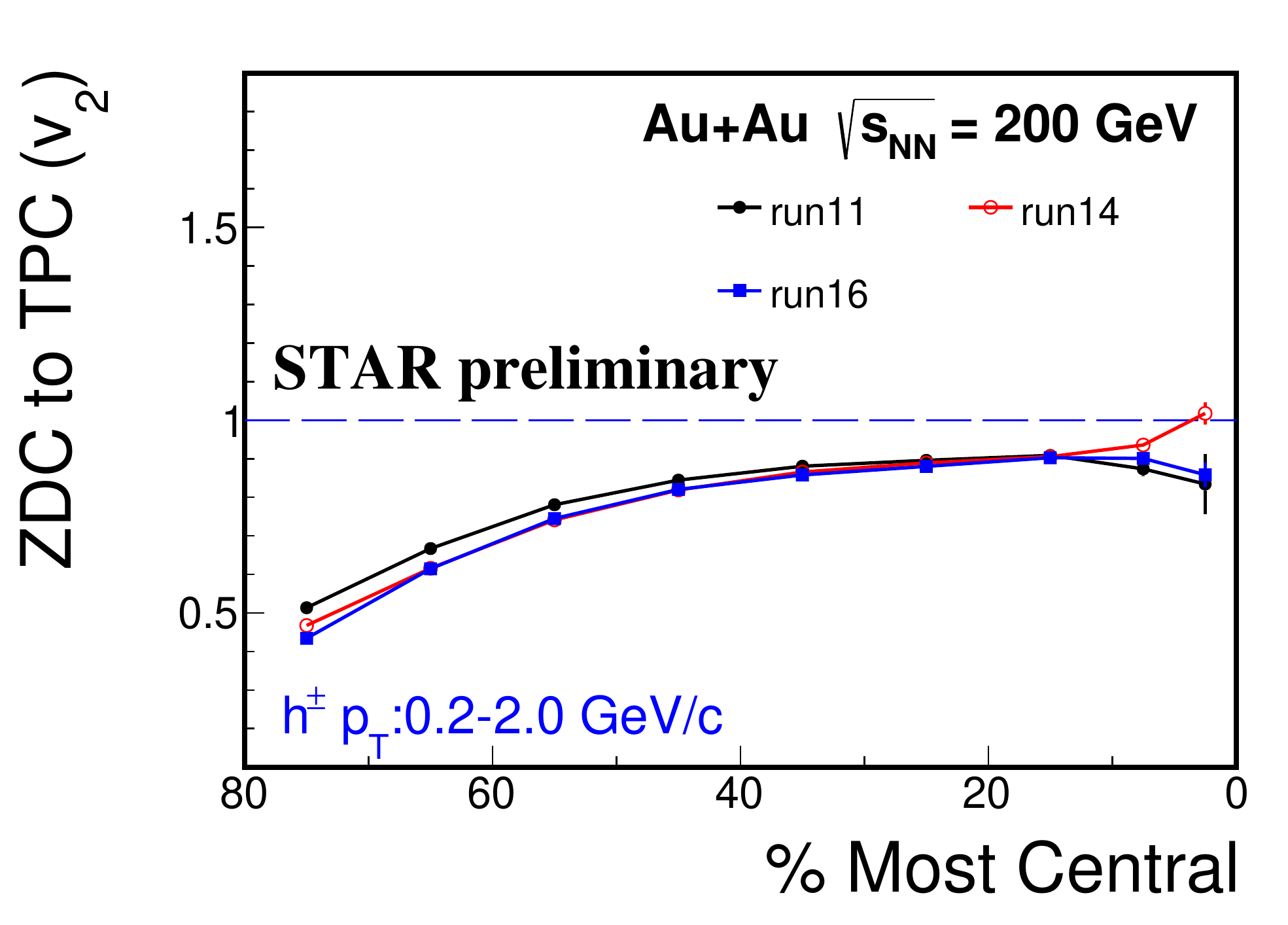}
  \includegraphics[width=0.4\textwidth]{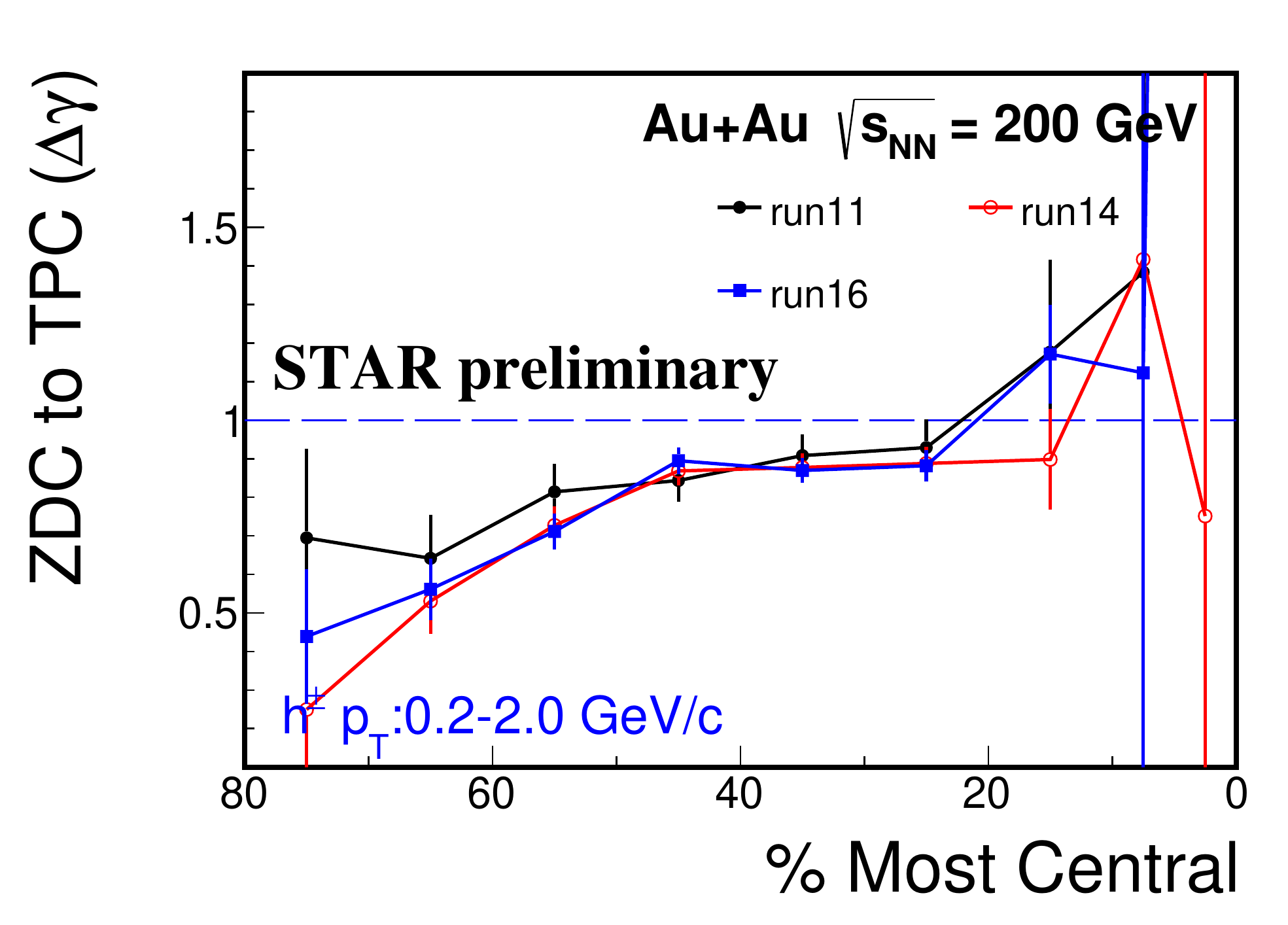}
  \includegraphics[width=0.4\textwidth]{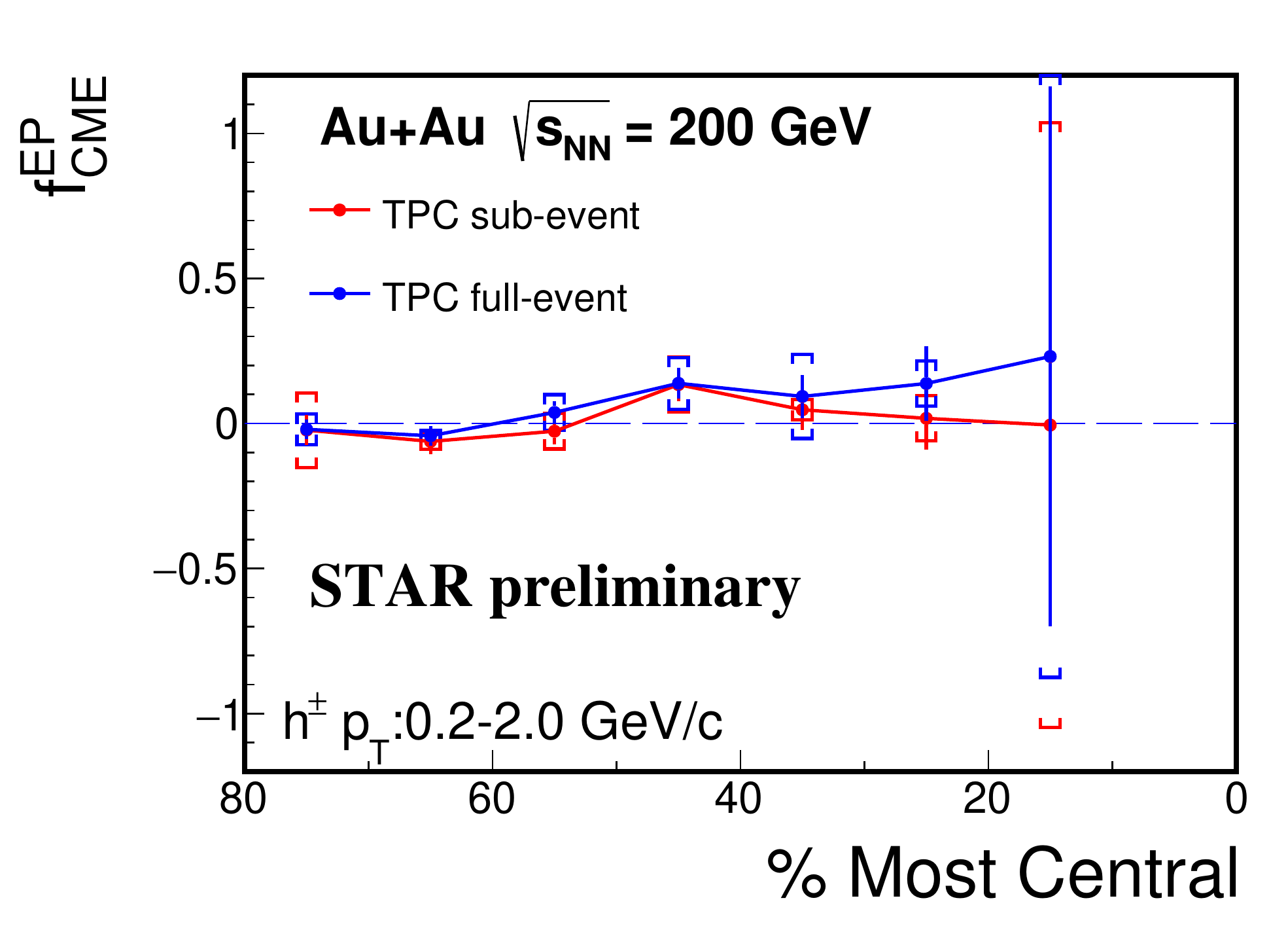}
  \caption{(Color online) The centrality dependences of the ratios of the $v_2$ (upper) and $\dg$ (middle) measured with respect to the ZDC event plane to those with respect to the TPC event plane. Lower: the extracted fraction of potential CME signal, $f_{\rm CME}^{\rm EP}$, as a function of collision centrality.}
  \label{fig:PPRP}
\end{figure}

STAR has employed this novel method to extract the CME signal~\cite{ZhaoQM18}.
Figure~\ref{fig:PPRP} upper panel shows the ratio of $v_2$ measured with respect to the ZDC 1st-order harmonic plane and that with respect to the TPC 2nd-order harmonic EP, and the middle panel shows the corresponding ratio of $\dg$~\cite{ZhaoQM18}. The sub-event method is used where the particles of interest ($\alpha$ and $\beta$) are from one half of the TPC in pseudorapidity and the reference particle ($c$) is from the other half. The lower panel of Fig.~\ref{fig:PPRP} shows the extracted CME fraction by Eq.~(\ref{EQ_RP10})~\cite{ZhaoQM18}. The full-event method $f_{\rm CME}^{\rm EP}$, where all three particles are from anywhere of the TPC, is also shown. Within errors, there is no measurable difference between sub-events and full events, though nonflow contribution is expected to be larger in the latter. 
The extracted CME fraction is $(9\pm4\pm7)$\% from TPC sub-events and $(12\pm4\pm11)$\% from TPC full events in 20-50\% centrality Au+Au collisions at 200~GeV~\cite{ZhaoQM18}.

\subsection{Invariant mass method}
\label{subsec:invmass}

It has been known since the very beginning that the $\dg$ were contaminated by background from resonance decays coupled with the elliptic flow ($v_{2}$)~\cite{Voloshin:2004vk,Wang:2009kd}; see Eq.~(\ref{EQ_2}).
Because of resonance elliptic anisotropy, more OS pairs align in the $\psiRP$ than the magnetic field direction, and it is an anti-charge separation along $\psiRP$.
This would mimic the same effect as the CME in the $\dg$ variable~\cite{Voloshin:2004vk,Abelev:2009ac,Abelev:2009ad}, which refers to the opposite-sign charges moving in the opposite directions along the magnetic field. 
Although the pair invariant mass ($\minv$) dependence of the $\dg$ would be the first thing to examine in terms of resonance background, it has been studied only recently~\cite{Zhao:2017nfq}.
The invariant mass provides the ability to identify and remove resonance decay backgrounds, enhancing the sensitivity of the $\dg$ measurements to potential CME signals.

\begin{figure}[htbp!]
  \centering 
  \includegraphics[width=0.42\textwidth]{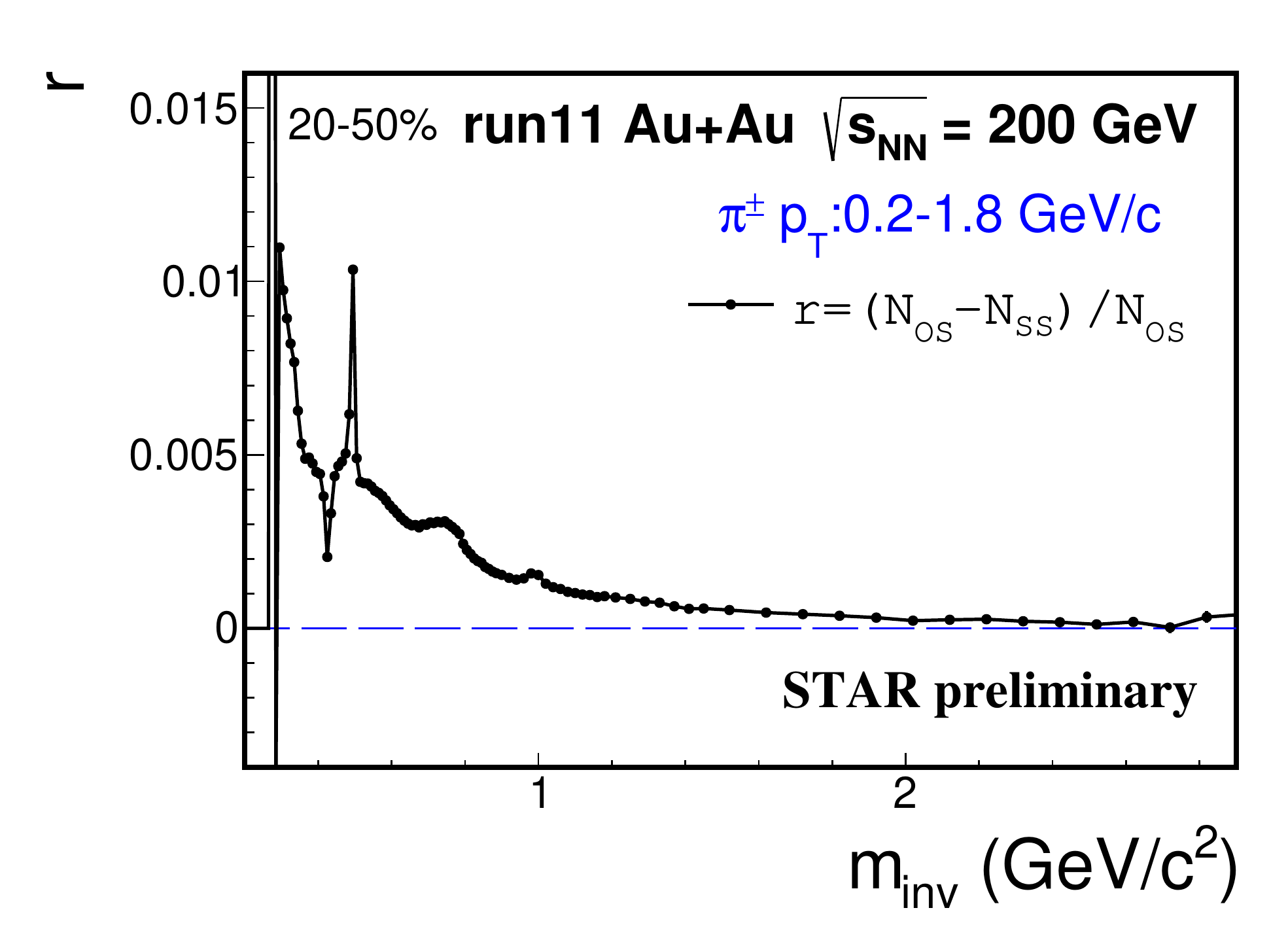}
  \includegraphics[width=0.42\textwidth]{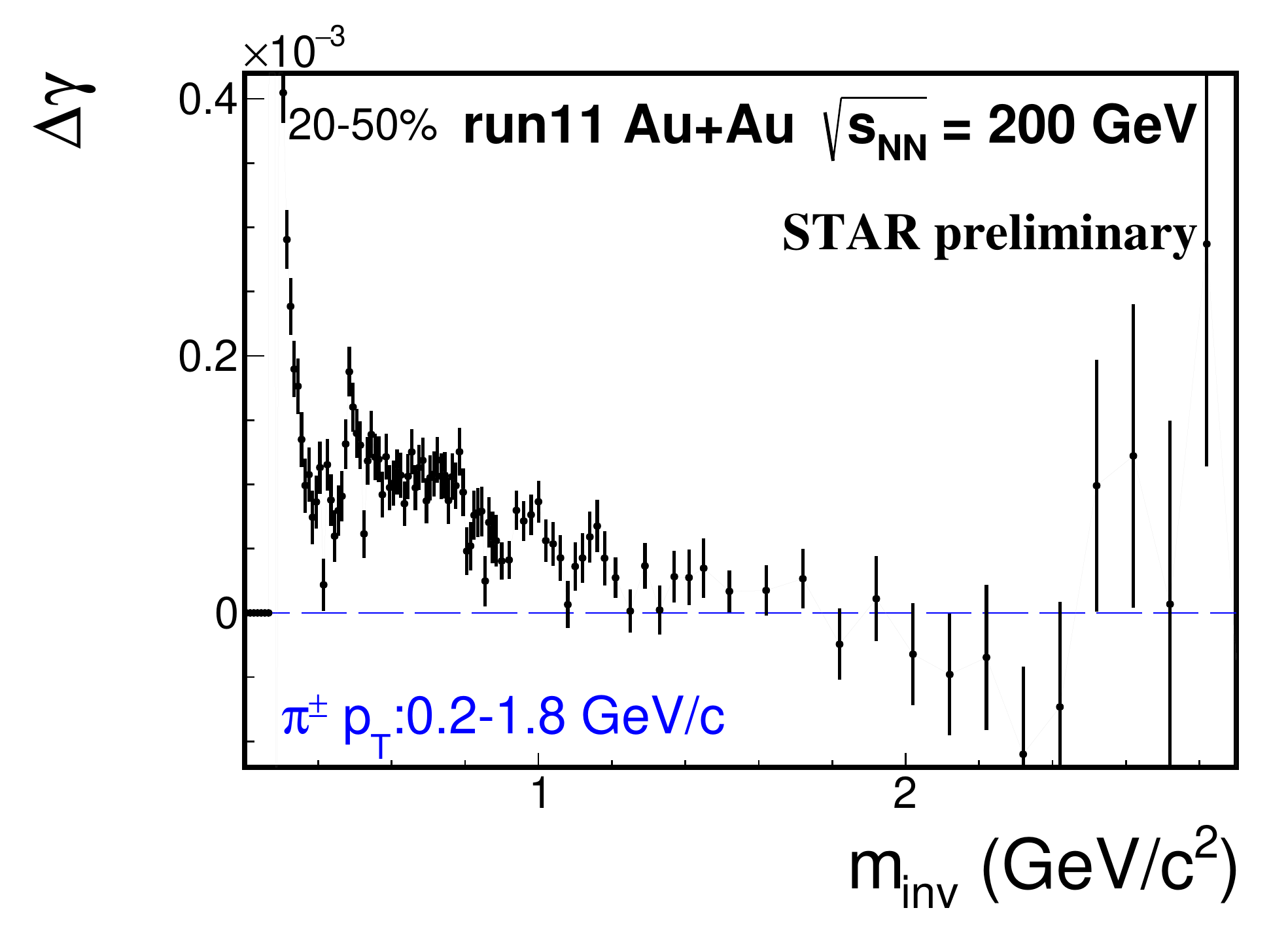}
  \caption{(Color online) The invariant mass ($\minv$) dependence of the relative excess of OS over SS pairs of charged pions (identified by the STAR TPC and TOF), $r=(N_{\rm OS}-N_{\rm SS})/N_{\rm OS}$ (upper), and the azimuthal correlator difference, $\dg=\gOS-\gSS$ (lower) in 20-50\% Au+Au collisions at \sNN = 200 GeV~\cite{Zhao:2017wck,ZhaoQM18}. Errors shown are statistical.} 
  \label{FG_IM2}
\end{figure}

Figure~\ref{FG_IM2} shows the preliminary results in mid-central Au+Au collisions from STAR experiments~\cite{Zhao:2017wck,ZhaoQM18}.
The upper panel shows the $\minv$ dependence of the relative OS and SS pair difference, $r=(N_{\rm OS}-N_{\rm SS})/N_{\rm OS}$; the lower panel shows that of the $\dg$ correlator. The $\minv$ structures are similar in $r$ and $\dg$. In other words, the $\dg$ correlator traces the distribution of the resonances. 

\begin{figure}
  \centering 
  \includegraphics[width=0.45\textwidth]{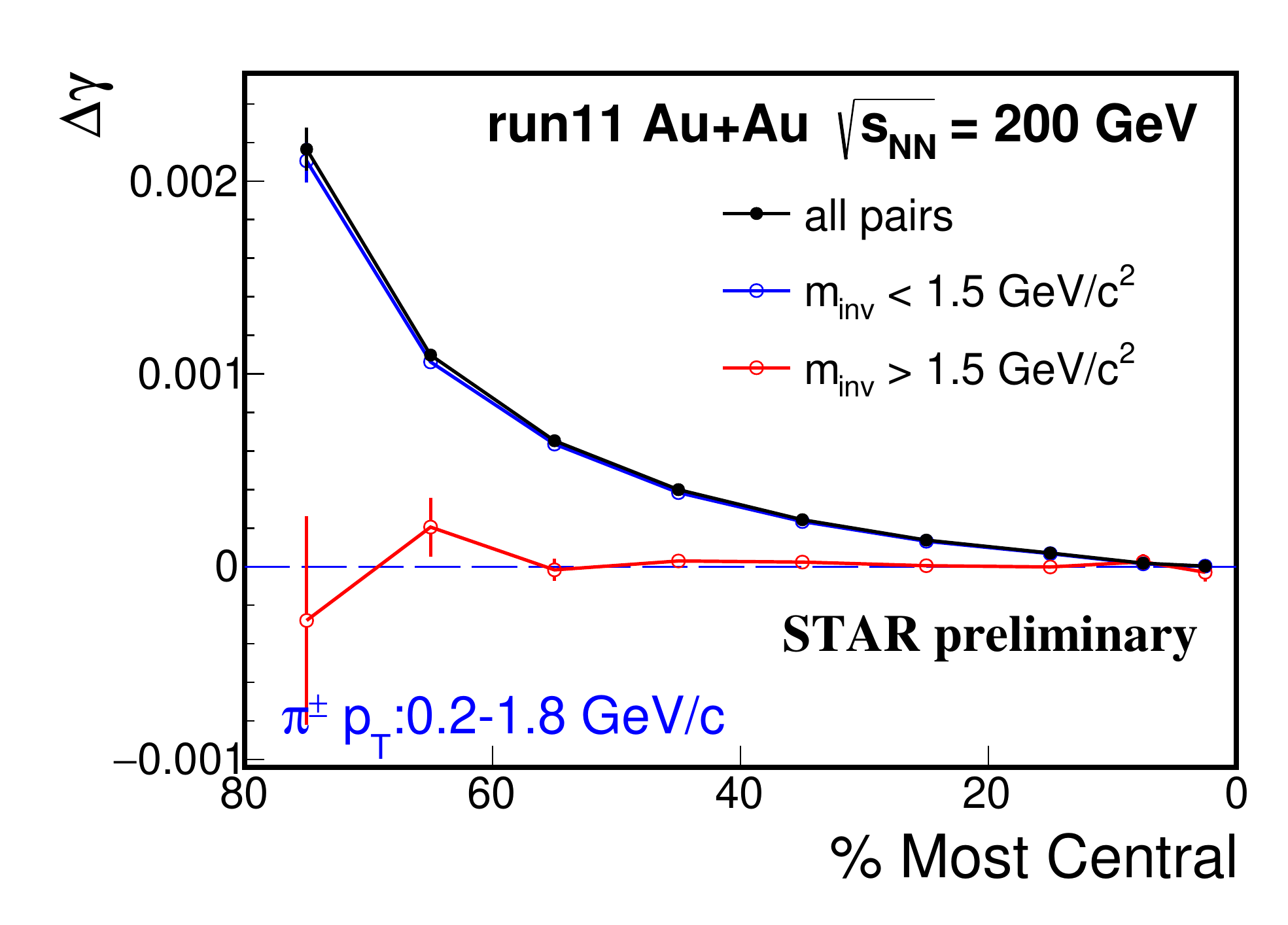}
  \caption{(Color online) The $\dg$ at $\minv > 1.5$ \gevcc (red) compared to the inclusive $\dg$ over the entire mass region (black) as a function of centrality in Au+Au collisions at 200 GeV~\cite{Zhao:2017wck,ZhaoQM18}.}
	\label{FG_IM1}
\end{figure}

Most of the $\pi$-$\pi$ resonances contributions are located in the low $\minv$ region~\cite{Agashe:2014kda,Adams:2003cc}.
It is possible to exclude them entirely by applying a lower $\minv$ cut.
Results from AMPT model show that such a $\minv$ cut, although significantly reducing the statistics, can eliminate essentially all resonance decay backgrounds~\cite{Zhao:2017nfq,Zhao:2017wck,ZhaoQM18}.
Figure~\ref{FG_IM1} shows the average $\dg$ with a lower mass cut, $\minv>1.5$~\gevcc, in comparison to the inclusive $\dg$ measurement~\cite{Zhao:2017wck,ZhaoQM18}. The high mass $\dg$ is drastically reduced from the inclusive data. Preliminary STAR data combining Run-11, 14, and 16 yield a $\dg$ at $\minv>1.5$~\gevcc\ of $(5\pm2\pm4)$\% of the inclusive $\dg$ measurements~\cite{ZhaoQM18}; the systematic uncertainty is currently estimated from the differences among the runs~\cite{ZhaoQM18}.

\begin{figure*}
  \centering 
  \includegraphics[width=0.7\textwidth]{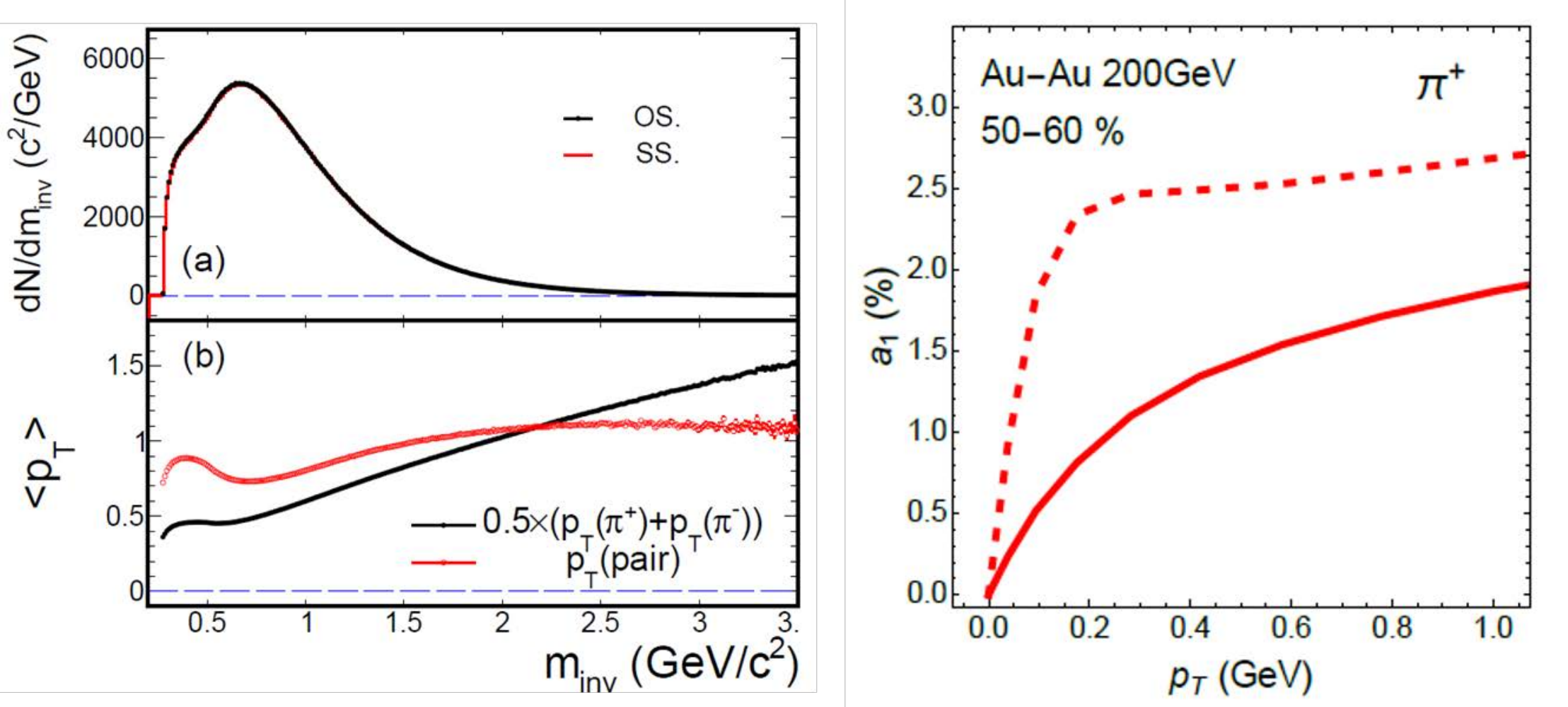}
  \caption{(Color online) Upper left: typical $\minv$ distributions of pion pairs in relativistic heavy-ion collisions. Lower left: the $\mean{\pt}$ of single pions (black) and of pion pairs (red) as functions of $\minv$~\cite{WangQM18}. Right: the CME charge separation signal strength in directly produced pions (dashed) and in final-state pions (solid) as functions of $\pt$~\cite{Shi:2017cpu}.}
  \label{fig:pt}
\end{figure*}

It is generally expected that the CME is a low $\pt$ phenomenon and its contribution to high mass may be small~\cite{Kharzeev:2007jp,Abelev:2009ad}. However, as shown in Fig.~\ref{fig:pt} left panel, a $\minv$ cut of 1.5~\gevcc\ corresponds to $\pt\sim 1$~\gevc\ which is not very high. 
Moreover, a recent study~\cite{Shi:2017cpu} indicates that the CME signal is rather independent of $\pt$ at $\pt>0.2$ \gevc\ (Fig.~\ref{fig:pt} right panel), suggesting that the signal may persist to high $\minv$.

Nevertheless, one can use the low $\minv$ data to extract the possible CME signal. In order to do so, resonance contributions must be subtracted.
In a two-component model, the $\minv$ dependence of the $\dg$ can be expressed~\cite{Zhao:2017nfq} as
\begin{equation}
  \dg(\minv) \approx r(\minv)R(\minv) + \dg_{CME}(\minv)\,. 
  \label{EQ_IM2}
\end{equation}
The first term is resonance contributions, where the response function 
$R(\minv)$ is a smooth function of $\minv$, while $r(\minv)$ contains resonance mass shapes. 
Consequently, the first term is not ``smooth'' but a peaked function of $\minv$.
The second term in Eq.~(\ref{EQ_IM2}) is the CME signal which should be a smooth function of $\minv$. 
The $\minv$ dependences of the CME signal and the background are distinctively different, and this can be exploited to identify CME signals at low $\minv$.
The feasibility of this method was investigated by a toy-MC simulation~\cite{Wang:2016iov} as well as in STAR data~\cite{Zhao:2017wck}. A linear response function $R(\minv)$ was assumed, guided by AMPT input~\cite{Wang:2016iov}, and various forms of CME$(\minv)$ were studied~\cite{Zhao:2017wck}.

\begin{figure}
  \centering 
  \includegraphics[width=0.48\textwidth]{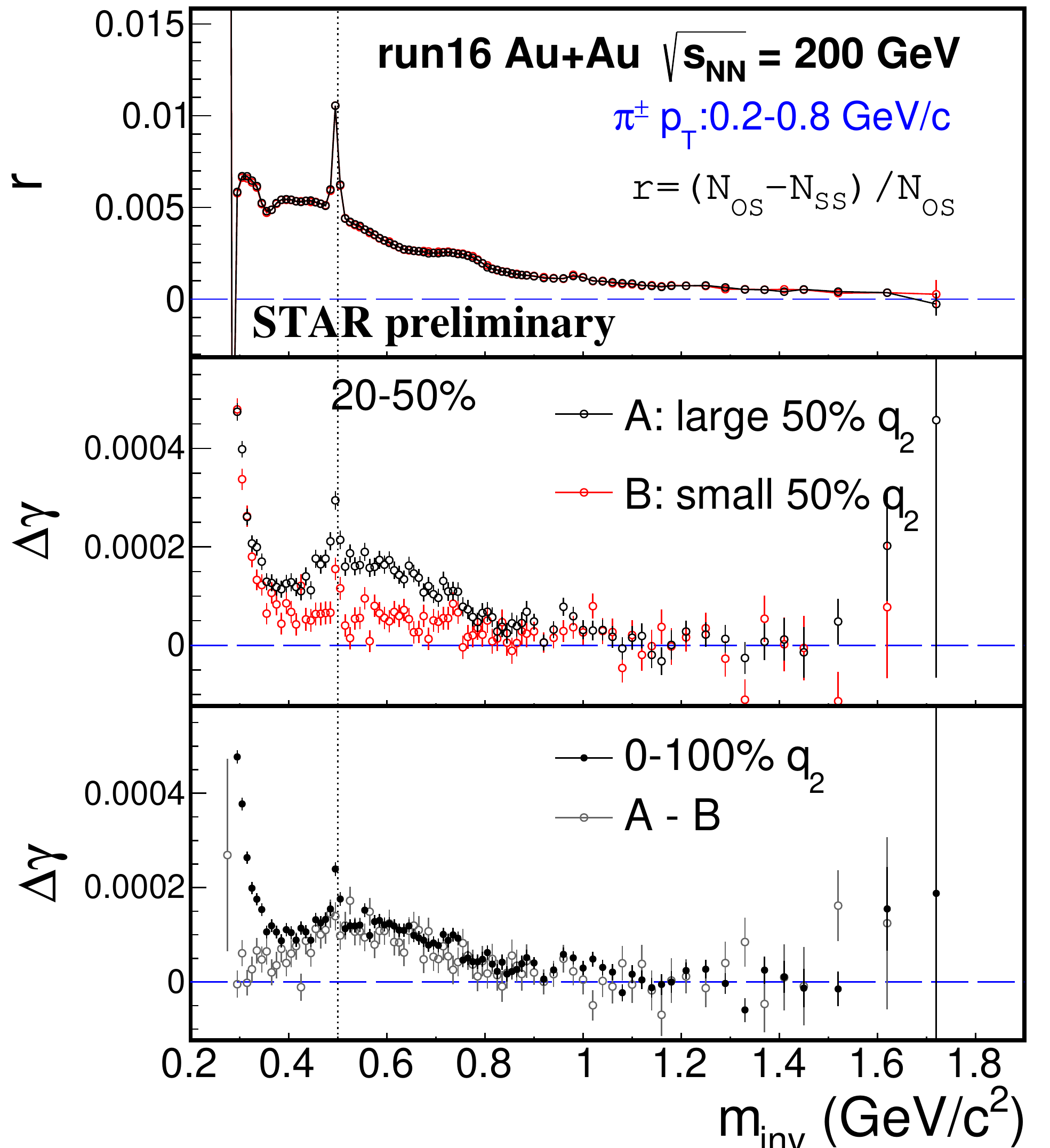}
  \caption{(Color online) The invariant mass ($\minv$) dependence of the relative excess of OS over SS pairs of charged pions (identified by the STAR TPC only), $r=(N_{\rm OS}-N_{\rm SS})/N_{\rm OS}$ (upper), the azimuthal correlator difference, $\dg=\gOS-\gSS$, of large and small $q_2$ events (middle), and the $\dg$ difference between large and small $q_2$ events together with the $\dg$ of all events (lower) in 20-50\% Au+Au collisions at \sNN = 200 GeV~\cite{ZhaoQM18}. Errors shown are statistical.}
  \label{fig:massESE}
\end{figure}

One difficulty in the above method is that the exact functional form of $R(\minv)$ is presently unknown and requires rigorous modeling and experimental inputs.
To overcome this difficulty, STAR has recently developed a method using the ESE technique~\cite{ZhaoQM18}. The events in each narrow centrality bin are divided into two classes according to the ExE $q_2$, calculated by Eqs.~(\ref{EQ_ESE1}) and~(\ref{EQ_ESE2}) using particles of interest. Since the magnetic fields are approximately equal for the two classes while the backgrounds differ, the difference in $\dg$ is a good representation of the background shape. Figure~\ref{fig:massESE} shows the $\dg(\minv)$ distributions for such two $q_2$ classes ($\dg_A$ and $\dg_B$) in the middle panel and the difference $\dg_A-\dg_B$ in the lower panel in 20-50\% Au+Au collisions~\cite{ZhaoQM18}. The $q_2$ selection is applied in narrower centrality bins than 20-50\%, and then the data are combined. The $\dg(\minv)$ of all events is also shown in the lower panel of Fig.~\ref{fig:massESE}. Note that the pion identification here was done using the TPC energy loss ($dE/dx$) information only, different from that in Fig.~\ref{FG_IM1}.

\begin{figure}
  \centering 
  \includegraphics[width=0.45\textwidth]{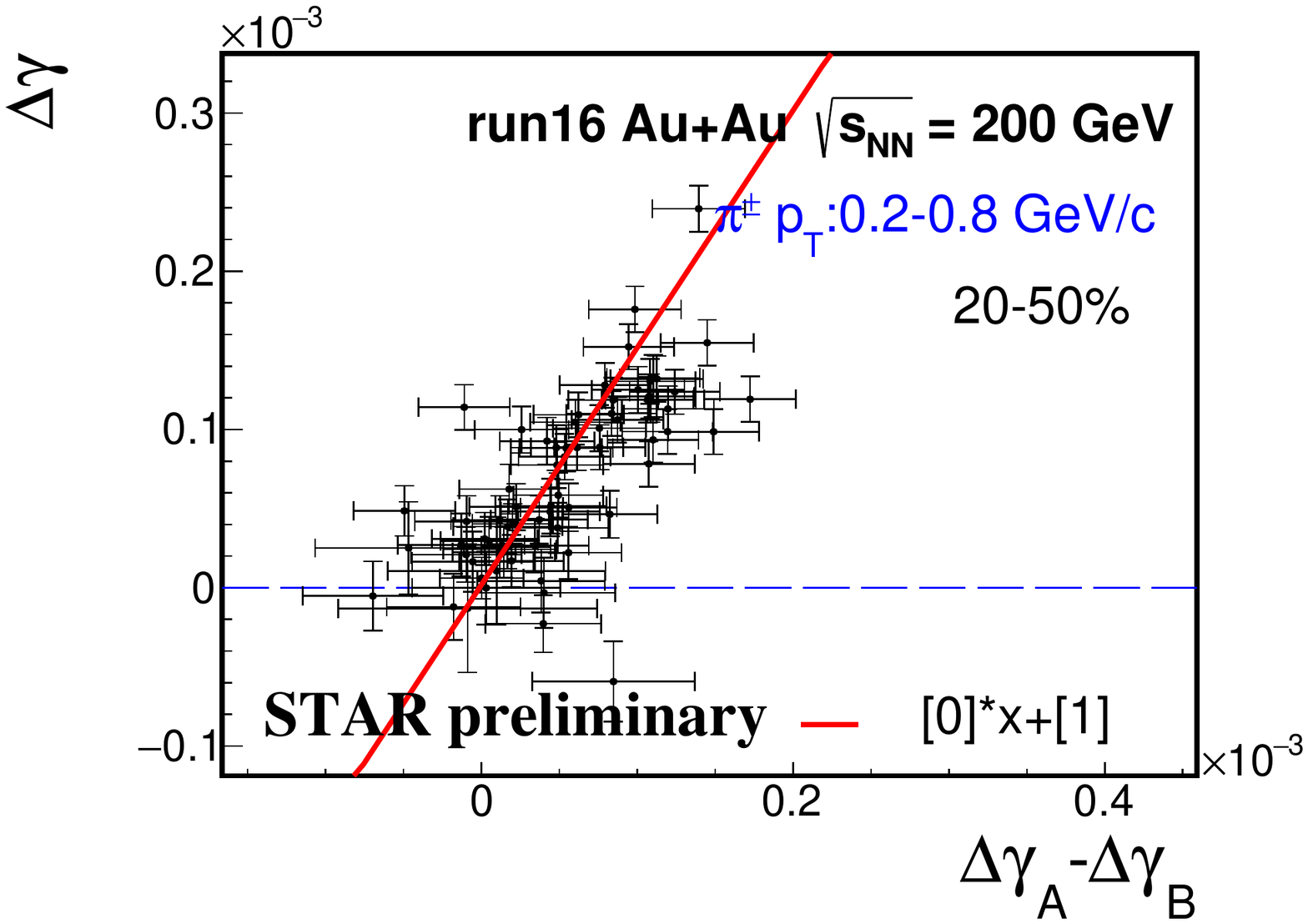}
  \includegraphics[width=0.45\textwidth]{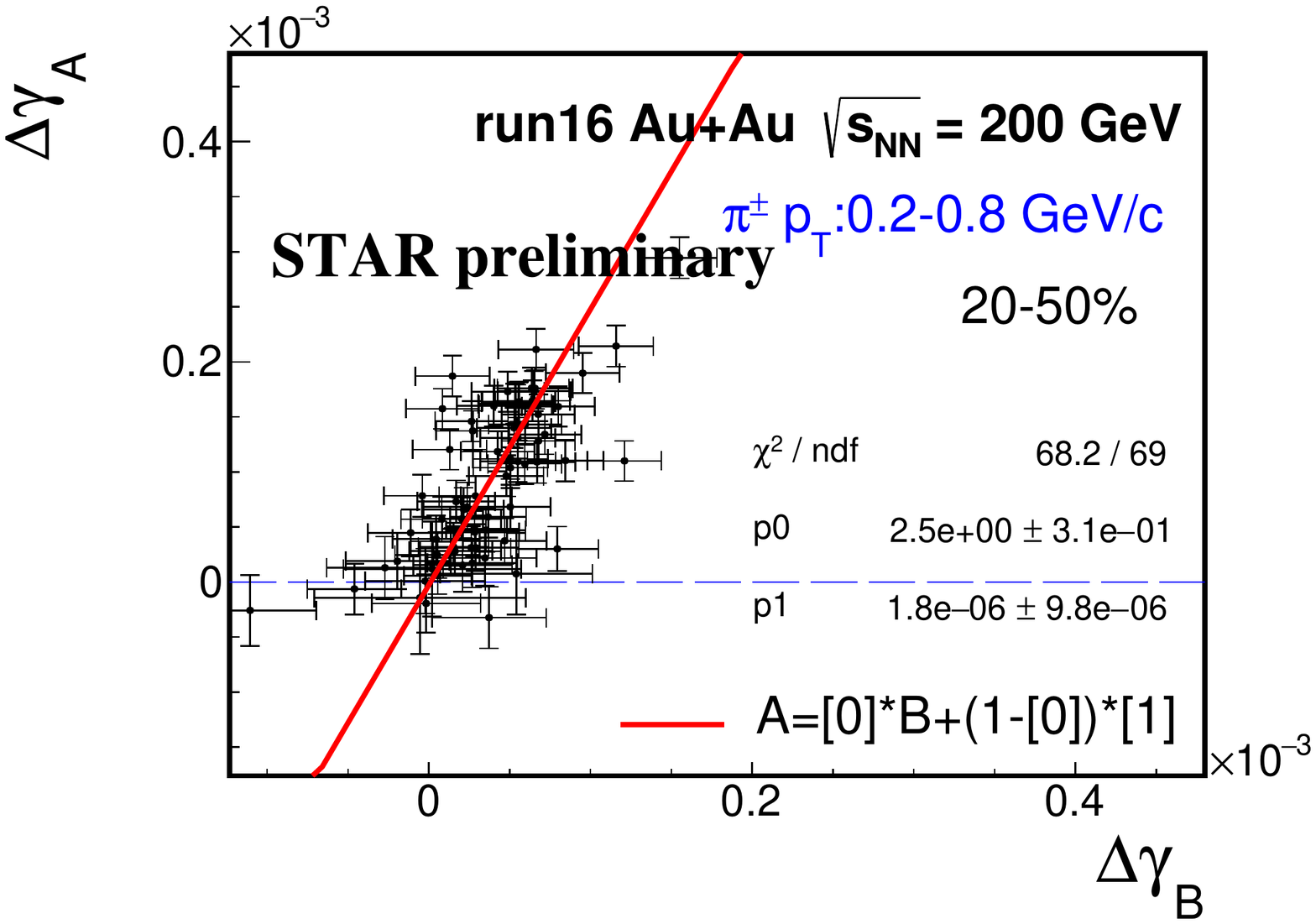}
  \caption{(Color online) $\dg$ versus $\dg_A-\dg_B$ (upper), and $\dg_A$ versus $\dg_B$ (lower) in 20-50\% centrality Au+Au collisions. Each data point corresponds to one $\minv$ bin in Fig.~\ref{fig:massESE}. Only the $\minv>0.4$~\gevcc\ data points are included.}
  \label{fig:massfit}
\end{figure}

The overall $\dg$ contains both background and the possible CME. With the background shape given by $\dg_A-\dg_B$, the CME can be extracted from a fit $\dg=k(\dg_A-\dg_B)+{\rm CME}$. Note that in this fit model the background is not required to be strictly proportional to $v_2$~\cite{WangQM18}. Figure~\ref{fig:massfit} upper panel shows $\dg$ as a function of $\dg_A-\dg_B$, where each data point corresponds to one $\minv$ bin in Fig~\ref{fig:massESE}~\cite{ZhaoQM18}. Only the $\minv>0.4$~\gevcc\ data points are included in Fig.~\ref{fig:massfit} because the $\dg$ from the lower $\minv$ region is affected by edge effects of the STAR TPC acceptance. Since the same data are used in $\dg$ and $\dg_A-\dg_B$, their statistical errors are somewhat correlated. To propoerly handle statistical errors, one can simply fit the indendent measurements of $\dg_A$ versus $\dg_B$, namely $\dg_A=b\dg_B+(1-b){\rm CME}$ where $b$ and ${\rm CME}$ are the fit parameters. Figure~\ref{fig:massfit} lower panel shows such a fit for the Run-16 Au+Au data~\cite{ZhaoQM18}.
Combining Run-11, 14, and 16 data, STAR obtained the possible CME signal to be $(2\pm4\pm6)$~\% of the inclusive $\dg$, where the systematic uncertainty is presently assessed from the differences among the runs~\cite{ZhaoQM18}.

\begin{figure}
  \centering 
  \includegraphics[width=0.45\textwidth]{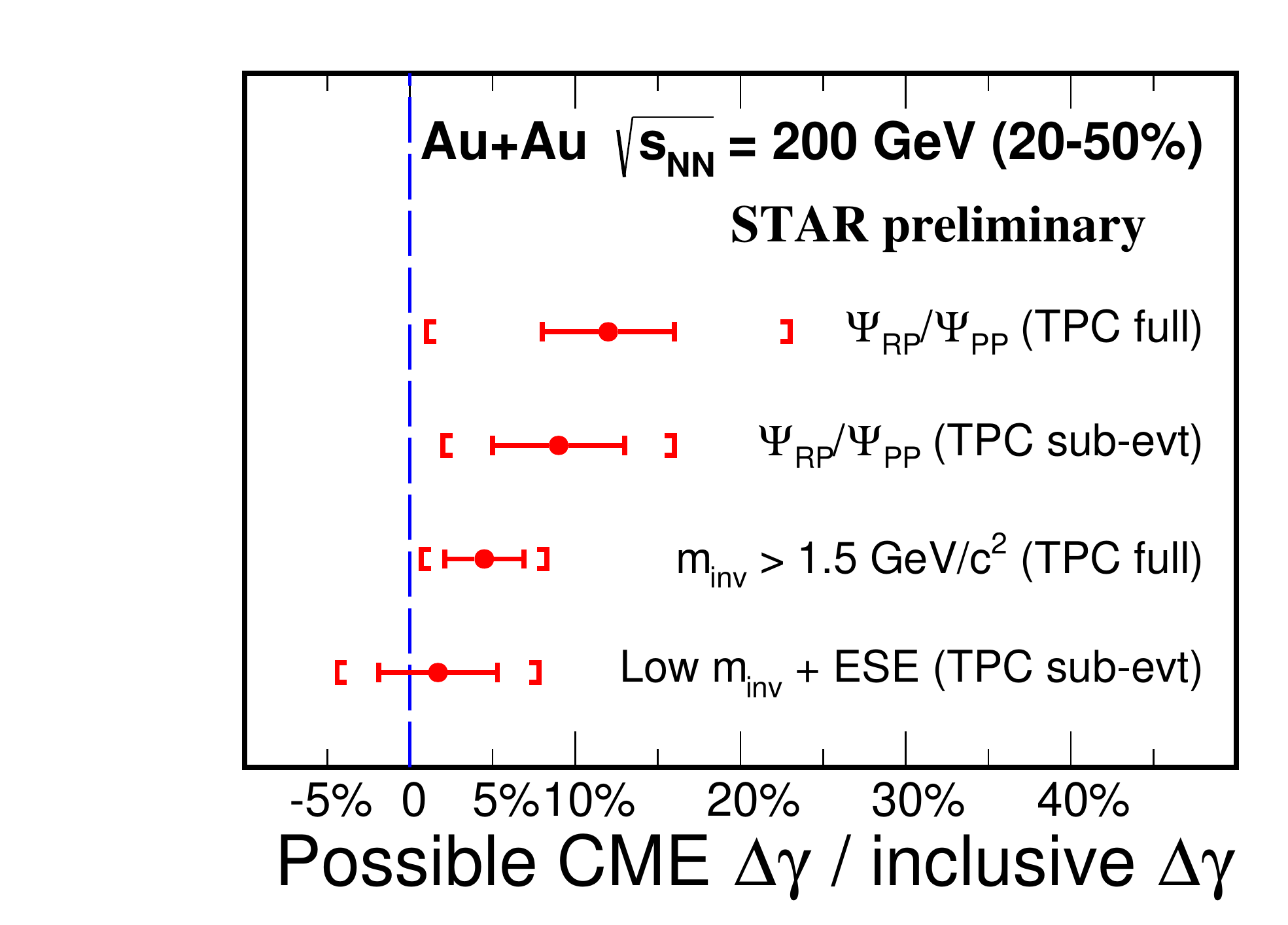}
  \caption{(Color online) The possible CME signal, relative to the inclusive $\dg$ measurement, extracted from the RP-PP comparative measurements and the invariant mass method, in 20-50\% centrality Au+Au collisions, with total 2.5 billion minimum-bias events combining Run-11 ($\sim$0.5B), Run-14 ($\sim$0.8B), and Run-16 ($\sim$1.2B).}
  \label{fig:QMfrac}
\end{figure}

Figure~\ref{fig:QMfrac} summarizes the current status of the CME results from STAR in 20-50\% centrality Au+Au collisions at $\snn=200$~GeV~\cite{ZhaoQM18}, using the novel methods described in this subsection and in Sect.~\ref{subsec:reactionplane}. The data~\cite{ZhaoQM18} show that the CME signal is small, on the order of a few percent of the inclusive $\dg$, with relatively large errors. Note that the data points in Fig.~\ref{fig:QMfrac} are from the same data using four different analysis methods. It is intriguing to note that all methods, although consistent with zero, seem to favor a positive value.
\section{Outlook}
\label{sec:outlook}

\begin{figure}[htbp!]
  \centering 
  \includegraphics[width=0.4\textwidth]{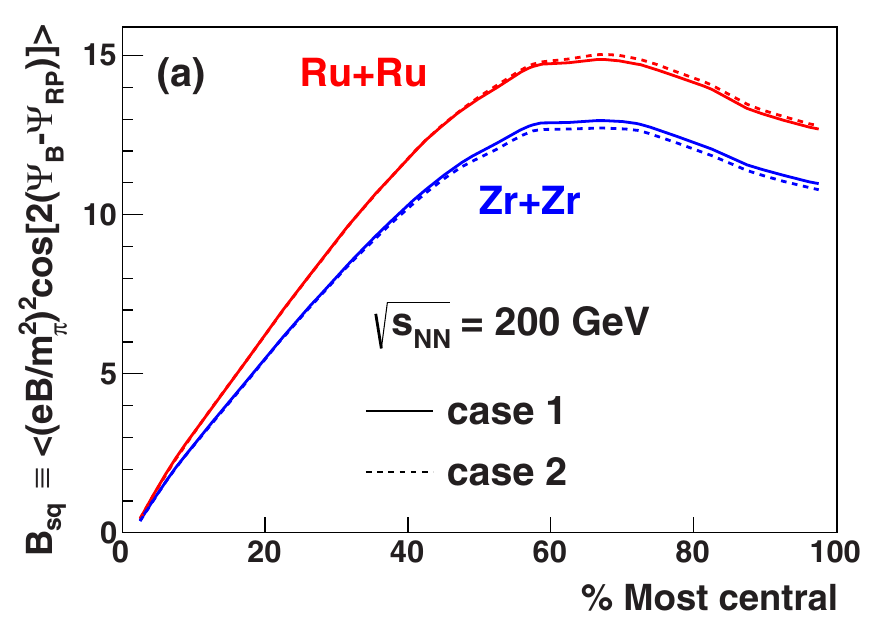} 
  \includegraphics[width=0.4\textwidth]{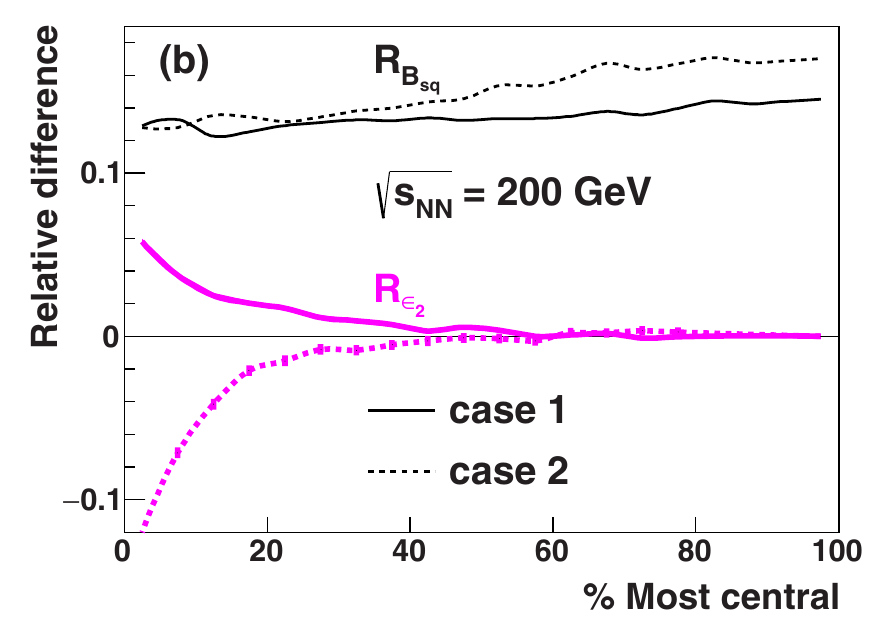} 
  \caption{(Color online) (a) Event-averaged initial magnetic field squared at the center of mass of the overlapping region, with correction from event-by-event fluctuations of the magnetic field azimuthal orientation, for Ru+Ru and Zr+Zr collisions at 200 GeV, and (b) their relative difference versus centrality. Also shown in (b) is the relative difference in the initial eccentricity. The line styles correspond to two extreme cases of the isobaric nuclear deformation parameters. From Ref.~\cite{Deng:2016knn}.}
  \label{FG_ISO1}
\end{figure}

The CME is related to the magnetic field while the background is produced by $v_{2}$-induced correlations. 
In order to gauge differently the magnetic field relative to the $v_{2}$, isobaric collisions have been proposed, such as $\RuRu$ and $\ZrZr$~\cite{Voloshin:2010ut}.
$\Ru$ and $\Zr$ have the same mass number but different charge (proton) number. One would thus expect the same $v_{2}$, which is insensitive to isospin, and 10\% difference in the magnetic field.
To test the idea of the isobaric collisions, MC Glauber calculations of the spatial eccentricity ($\epsilon_{2}$) and the magnetic field strength in Ru+Ru and Zr+Zr collisions have been carried out~\cite{Deng:2016knn,Huang:2017azw}. 
The Woods-Saxon spatial distribution is used~\cite{Deng:2016knn,Huang:2017azw},
\begin{equation}
  \rho(r,\theta) = \frac{\rho_0}{1+\rm{exp}\{[r-R_{0}-\beta_{2}R_{0}Y_{2}^{0}(\theta)]/ a \}}\,,
  \label{EQ_ISO1}
\end{equation}
where $R_{0}$ is the charge radius parameter of the nucleus, $a$ represent the surface diffuseness parameter, $Y_{2}^{0}$ is the spherical harmonic, and $\rho_{0}$ is the normalization factor. The parameter $a\approx0.46$~fm is almost identical for $\Ru$ and $\Zr$.  The charge radii of $R_{0}=5.085$~fm and 5.020~fm were used for $\Ru$ and $\Zr$, respectively, for both the proton and neutron densities.
The deformity quadrupole parameter $\beta_{2}$ has large uncertainties; extreme cases were taken and yielded less than 2\% difference in $\epsilon_{2}$ between Ru+Ru and Zr+Zr collisions in the 20-60\% centrality range~\cite{Deng:2016knn,Huang:2017azw}.
The magnetic field strengths in Ru+Ru and Zr+Zr collisions were calculated by using Lienard-Wiechert potentials with the HIJING model taking into account the event-by-event azimuthal fluctuations of the magnetic field orientation~\cite{Deng:2012pc}. The quantity relevant to the CME is the average magnetic field squared with correction from the event-by-event azimuthal fluctuation of the magnetic field orientation,  
\begin{equation}
  \Bsq \equiv \mean{(eB/m_{\pi}^{2})^{2} \cos[2(\psi_{B} - \psiRP)]}\,.
\end{equation}
Figure~\ref{FG_ISO1}(a) shows the calculated $\Bsq$ at the initial encounter time of the nuclei in Ru+Ru and Zr+Zr collisions at 200 GeV. 
Figure~\ref{FG_ISO1}(b) shows the relative difference in $\Bsq$,
\begin{equation}
  R_{\Bsq}=2(\Bsq^{\rm Ru+Ru}-\Bsq^{\rm Zr+Zr})/(\Bsq^{\rm Ru+Ru}+\Bsq^{\rm Zr+Zr})\,.
\end{equation}
The difference is approximately 15\%. 
Figure~\ref{FG_ISO1}(b) also shows the relative difference in the initial eccentricity,
\begin{equation}
  R_{\epsilon_{2}}=2(\epsilon_{2}^{Ru+Ru}-\epsilon_{2}^{Zr+Zr})/(\epsilon_{2}^{Ru+Ru}+\epsilon_{2}^{Zr+Zr})\,.
\end{equation}
The relative difference in $\epsilon_{2}$ is practically zero, at most 2\% in 20-60\% centrality. 
This suggests that the $v_{2}$-induced backgrounds are almost the same for Ru+Ru and Zr+Zr collisions in the 20-60\% centrality range. 

Based on the available experimental $\dg$ measurements in Au+Au collisions at 200 GeV and the calculated magnetic field and eccentricity differences between Ru+Ru and Zr+Zr collisions, it was estimated that 400 million events each for Ru+Ru and Zr+Zr collisions, assuming 1/3 of the currently measured $\dg$ to be CME signal, would yield a $5\sigma$ difference between the two systems~\cite{Deng:2016knn,Huang:2017azw}. The isobar run, just concluded at RHIC, has accumulated 2 billion events each for Ru+Ru and Zr+Zr collisions in the STAR detector. If the CME signal is 5\% of the inclusive $\dg$ measurement, as implied by the latest STAR results~\cite{ZhaoQM18}, then the isobar data would yield a 1-2$\sigma$ effect.

\begin{figure}[htbp!]
  \centering 
  \includegraphics[width=0.35\textwidth]{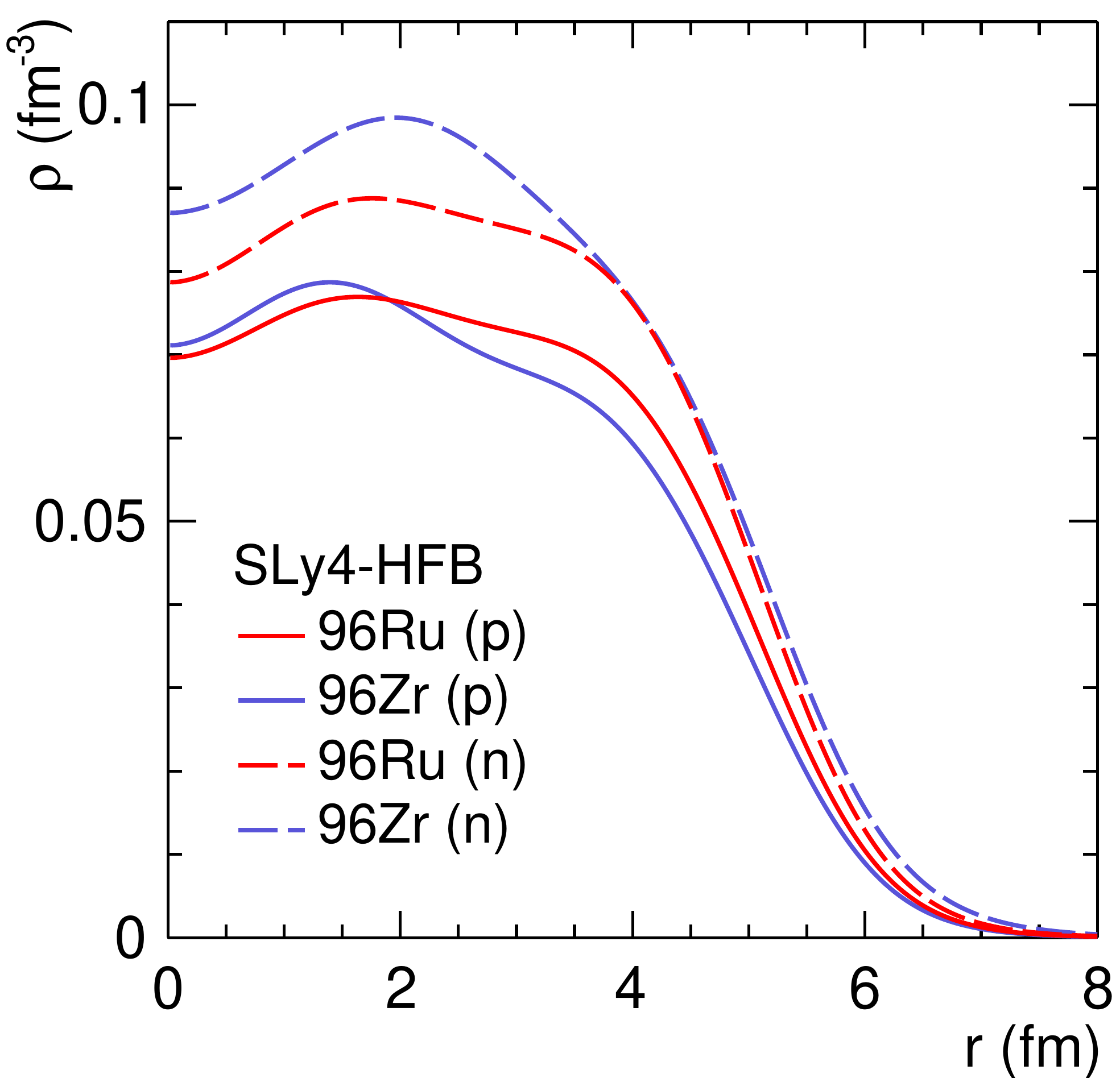} 
  \includegraphics[width=0.36\textwidth]{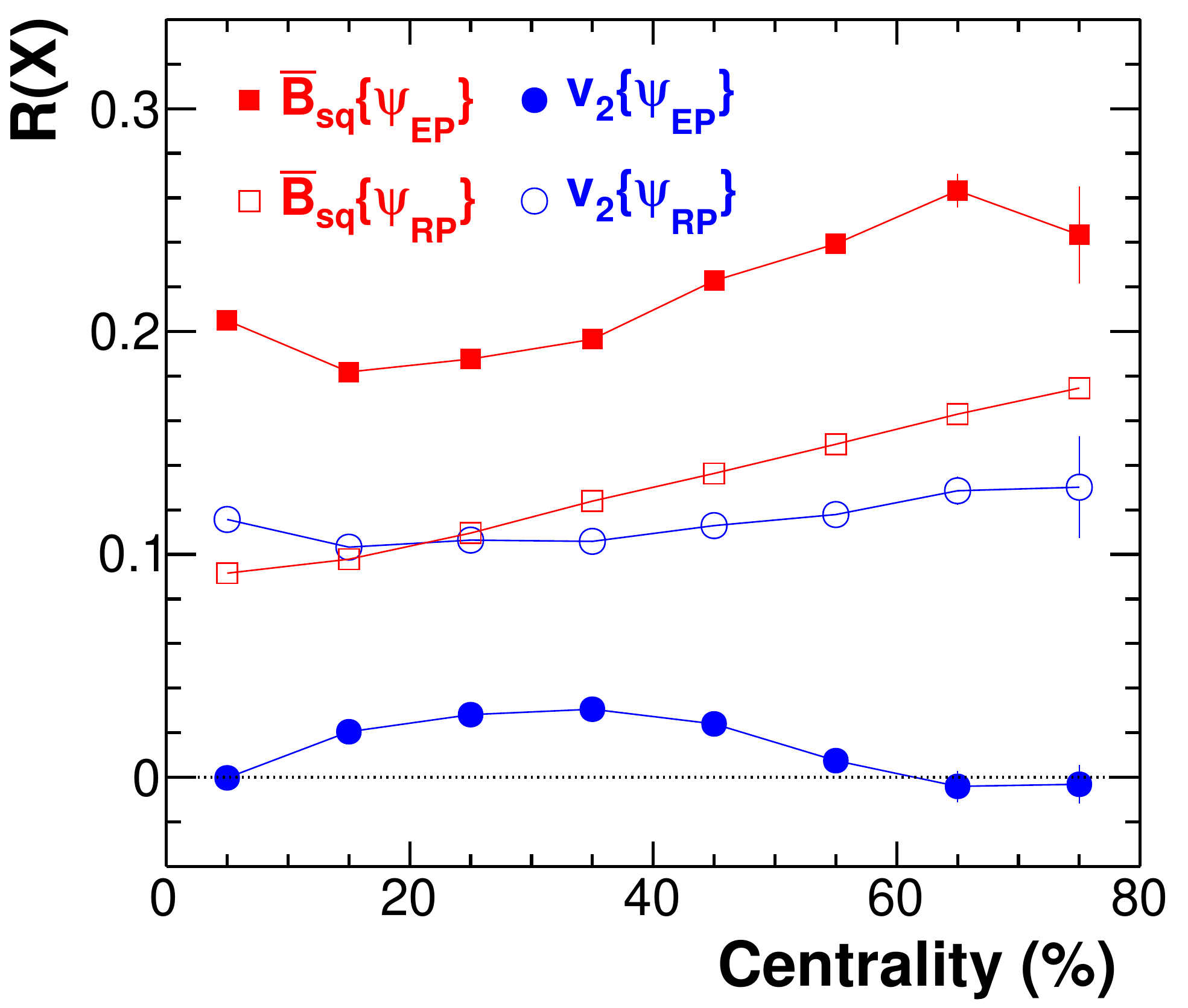} 
  \caption{(Color online) Upper: proton and neutron density distributions of the $\Ru$ and $\Zr$ nuclei, assumed spherical, calculated by DFT~\cite{Xu:2017zcn}. Lower: relative differences between Ru+Ru and Zr+Zr collisions as functions of centrality in $v_{2}\{\psi\}$ and $\Bsq\{\psi\}$ with respect to $\psiRP$ and $\psiEP$ from AMPT simulations using the DFT densities from the upper panel~\cite{Xu:2017zcn}.}
  \label{FG_ISO3}
\end{figure}
The above estimates assume Woods-Saxon densities, identical for proton and neutron distributions.
Using the energy density functional theory (DFT) with the well-known SLy4 mean field~\cite{Chabanat:1997un} including pairing correlations (Hartree-Fock-Bogoliubov, HFB approach)~\cite{Wang:2016rqh,Bender:2003jk,ring1980nuclear}, the ground-state density distributions of $\Ru$\ and $\Zr$, assumed spherical, were calculated~\cite{Xu:2017zcn}. The results are shown in the upper panel of Fig.~\ref{FG_ISO3}~\cite{Xu:2017zcn}. 
They show that protons in Zr are more concentrated in the core, while protons in Ru, 10\% more than in Zr, are pushed more toward outer regions. 
The neutrons in Zr, four more than in Ru, are more concentrated in the core but also more populated on the nuclear skin.
The lower panel of Fig.~\ref{FG_ISO3} shows the relative differences in $v_{2}\{\psi\}$ and $\Bsq\{\psi\}$ between Ru+Ru and Zr+Zr collisions as functions of centrality from AMPT simulations with the densities calculated by the DFT method~\cite{Xu:2017zcn}. Results with respect to both $\psiRP$ and $\psiEP$ are depicted. They suggest that the relative difference in $\epsilon_{2}$ and $v_{2}$ with respect to $\psiEP$ are as large as $\sim$3\%, and that in $\Bsq$ is the expected $\sim$20\%. With respect to $\psiRP$, the differences in $v_{2}$ and $\Bsq$ are both on the order of 10\%. 
These results suggest that the premise of isobaric sollisions for the CME search may not be as good as originally anticipated, and could provide important guidance to the experimental isobaric collision program.

No matter what the outcome of the isobaric collision data is, the search for the CME shall continue. More statistics should be accumulated for Au+Au collisions at RHIC and Pb+Pb collisions at the LHC. Future detector upgrades should be considered to improve the sensitivities to the CME. Additional novel analysis techniques should be developed.
\section{summary}
\label{sec:summary}

Relativistic heavy-ion collisions provide an ideal environment to study the the chiral magnetic effect (CME) induced by topological charge fluctuations in QCD. 
Since the first three-point correlator ($\gamma$) measurements in 2009, experimental results have been abundant in relativistic heavy-ion as well as small system collisions.
Those measurements are contaminated by major physics backgrounds. 
In this article, experimental efforts in addressing those backgrounds in both heavy-ion and small-system collisions are reviewed, and several novel methods to search for the CME with various background sensitivities are discussed. These include event-by-event elliptic flow ($v_2$), event-shape engineering, comparative measurements with respect to the participant plane (PP) and reaction plane (RP), and pair invariant-mass ($\minv$) dependence. 
The current estimates on the strength of the possible CME signal are on the order of a few percent of the inclusive $\dg$ values, consistent with zero with large uncertainties. The prospect of the recently taken isobaric collision data is discussed.

It is clear that the experimental challenges in the CME search are dauting. Major efforts have been devoted to the CME search from both experimental and theoretical sides (the latter is not reviewed here). There is no doubt that the physics of the CME is of paramount importance. The unremitting pursuit for the CME in heavy-ion collisions will not be wasted. 
\begin{acknowledgments}
  We thank Professor Wei Li for valuable discussions. J.Z.~and F.W.~thank Professor Zi-Wei Lin, Professor Hanlin Li, and Mr.~Yicheng Feng for collaboration.
  The work of J.Z.~and F.W.~was supported in part by National Natural Science Foundation of China (Grant No.~11747312) and U.S.~Department of Energy (Grant No.~de-sc0012910).
  The work of Z.T.~was supported in part by an Early Career Award (Contract No.~de-sc0012185) from the U.S.~Department of Energy Office of Science, the Robert Welch Foundation (Grant No.~C-1845) and an Alfred P. Sloan Research Fellowship (No.~FR-2015-65911).
\end{acknowledgments}

\bibliography{ref}
\end{document}